\documentclass[a4paper,11pt,preprintnumbers]{article}

\usepackage{fullpage}

\usepackage[british]{babel}
\usepackage{verbatim}
\usepackage[T1]{fontenc}
\usepackage{lmodern}
\usepackage{lipsum}
\usepackage{booktabs}
\usepackage{caption}

\usepackage{amsthm, amssymb}
\usepackage[tbtags]{amsmath}
\usepackage{mathtools}
\usepackage{amstext}
\usepackage{braket}
\usepackage{multirow}

\usepackage{hyperref}

\numberwithin{equation}{section}

\begin{document} 
	
\begin{titlepage}
\thispagestyle{empty}
\begin{flushright}
	\hfill{DFPD-2016/TH/10} 
\end{flushright}
				
\vspace{35pt}
				
\begin{center}
	{\LARGE{\bf Interactions of ${\cal N}$ Goldstini in Superspace}}
									
	\vspace{50pt}
							
	{N.~Cribiori, G.~Dall'Agata and	 F.~Farakos}
							
	\vspace{25pt}
							
	{
		{\it  Dipartimento di Fisica ``Galileo Galilei''\\
			Universit\`a di Padova, Via Marzolo 8, 35131 Padova, Italy}
										
		\vspace{15pt}
										
		{\it   INFN, Sezione di Padova \\
		Via Marzolo 8, 35131 Padova, Italy}
		}
								
\vspace{40pt}
								
{ABSTRACT} 
\end{center}
				
We study field theories with ${\cal N}$ extended non-linearly realized supersymmetries, describing the couplings of models that contain ${\cal N}$ goldstini.
We review all the known formulations of the ${\cal N}=1$ goldstino theories and we generalize them to an arbitrary number ${\cal N}$ of non-linearly realized supersymmetries.
We explicitly prove the equivalence of all these extended supersymmetry breaking models containing ${\cal N}$ goldstini and reformulate the theory with ${\cal N}$ supersymmetries in terms of standard ${\cal N}=1$ constrained superfields.

\vspace{10pt}
			
\bigskip
			
\end{titlepage}

\baselineskip 6 mm

\tableofcontents

\section{Introduction} 
\label{sec:introduction}

One of the first appearances of supersymmetry was to describe the seemingly massless neutrino as a goldstone particle whose interactions were dictated by a non-linear symmetry acting as a fermionic shift \cite{Volkov:1973ix}.
In the meanwhile supersymmetry was also formulated as a linear symmetry that could address the gauge hierarchy problem.
This fact triggered a systematic use of supersymmetry in Particle Physics and led to the supersymmetrization of the Standard Model (see \cite{Martin:1997ns} for a review).
In this framework non-linearly realized supersymmetry can serve as an organizing principle for the description 
of low energy effective theories, because it encodes information of the underlying linearly realized ones.
As an example, it can be used to introduce soft breaking terms in the Minimal Supersymmetric Standard Model \cite{Antoniadis:2010hs}.

In most applications the global non-linear supersymmetry is exact and the fermionic goldstone modes are massless.
A proposal conceptually closer to the original motivation underlying the work of Volkov--Akulov, however, is to study light fermions as pseudo-goldostone modes of an approximate supersymmetry.
In this respect supersymmetry remains non-linearly realized and it does not necessarily have an UV completion in terms of a linearly realized representation.
Strongly coupled sectors, in particular, can have such pseudo-goldstone modes.
As an example, there are models that describe quarks and leptons of the Standard Model as remnants of some of these strongly coupled sectors \cite{Bardeen:1981df,Liu:2016idz}.
In this scenario, softly violated non-linear realizations of supersymmetry can be used as a tool for studying the low energy effective theory associated to some strongly coupled sector.

Superspace is the natural framework for the formulation of supersymmetric theories \cite{Gates:1983nr,Wess:1992cp}.
The development of a superspace formalism for ${\cal N}$ extended non-linearly realized supersymmetries is therefore compelling.
Geometric methods for studying this type of theories have been defined \cite{Clark:2000rv}, but a complete superspace setup has not been established yet, even though scattered results do exist \cite{Samuel:1982uh,Ferrara:1983fi,Kandelakis:1986bj}.
Our work aims exactly to set the foundations for this program.
The starting point is the supersymmetry breaking sector.
It contains  
${\cal N}$ goldstini $G_I$, with $I=1, \dots {\cal N}$, and the auxiliary field ${\cal F}$ giving the supersymmetry breaking scale.
Once the properties of this sector are established, additional matter and gauge constrained superfields can be coupled to it \cite{Rocek:1978nb}-\cite{Dall'Agata:2016yof}.
Considering ${\cal N}>4$ supersymmetries, for example, some component fields of the goldstini supermultiplet are going to have spin higher than one.
Due to the non-linear realization, these fields are removed in terms of the ${\cal N}$ goldstini, much in the same way as the sgoldstino is removed from the spectrum in the ${\cal N}=1$ theory.

Our main result is the construction of the supersymmetry breaking sector for a generic number ${\cal N}$ of supersymmetries, in the case they are all spontaneously broken and within different superspace formulations.
We study first the system of ${\cal N}$ goldstini in the Samuel--Wess formalism \cite{Samuel:1982uh,Kandelakis:1986bj} and we prove the equivalence between this formalism and the geometric method \cite{Volkov:1973ix,Clark:2000rv}.
We then present the ${\cal N}$ goldstini model in the somehow generalized formalism of \cite{Rocek:1978nb,Lindstrom:1979kq}, identifying the generalization of Rocek's constraints to ${\cal N}$ supersymmetries.
In a more modern approach, we also reformulate the Lagrangian and interactions of ${\cal N}$ goldstini 
in terms of constrained superfields \cite{Casalbuoni:1988xh,Komargodski:2009rz}.
This can be done by considering a chiral superfield in ${\cal N}$ superspace 
\begin{equation}
	\nonumber
	\bar D_{\dot \alpha J}{\cal X} = 0\,, 
\end{equation}
and imposing the constraints   
\begin{equation}
	\nonumber
	{\cal X} (D^{I\neq J})^{2\mathcal{N}-2}D^J_\alpha  {\cal X} =0\,,
\end{equation}
which imply also 
\begin{equation} 
	\nonumber
	{\cal X}^2 =0\,.
\end{equation}
These constraints remove all component fields from the spectrum except the goldstini and the auxiliary field acquiring the vacuum expectation value.
We express finally our results in terms of ${\cal N}=1$ constrained superfields.
In particular, we relate the discussion to the standard nilpotent superfield methods and to the  chiral superfields system \cite{Brignole:1997pe,Komargodski:2009rz} 
\begin{equation}
	\nonumber
	X^2 =0, \qquad  X Y_i =0, \qquad \bar{X} D^2 Y_i = 0 ,
\end{equation}
where the ${\cal N}=1$ goldstino is accomodated inside $X$, while the other goldstini are inside $Y_i$.
The couplings of these superfields are dominated by the additional non-linearly realized supersymmetries.
We briefly present also an alternative way to describe the theory, accomodating the goldstini inside the chiral superfields $X$ and $H^{i}_{\dot{\alpha}}$ such that
\begin{equation}
	\nonumber
	X^2 =0, \qquad  \bar D_{\dot \alpha} (X \bar H_{i \beta} ) =0 .
\end{equation}

The possible applications of the formalism we developed are numerous.
Let us therefore sketch a few of them before concluding this introductory section.
One first direction might be the study of theories where ${\cal N}$ supersymmetries have been spontaneously broken at some high energy scale $f$ and they become non-linearly realized.
In particular the possible matter couplings can be investigated, as for example in \cite{Antoniadis:2010hs}, and it would be interesting to understand the restrictions on these couplings imposed by the low energy theory.
In fact one could even go up to ${\cal N}=8$ supersymmetries, since it is known that gravitino interactions are dominated by the spin-1/2 part during high energy scatterings \cite{Casalbuoni:1988kv}.
In these models one is assuming the UV theory to have a linear realization of supersymmetry.
An alternative direction is to assume the UV theory is in fact a strongly coupled sector and therefore the non-linear supersymmetry is not exact.
The goldstini are then massive pseudo-goldstone modes.
This is an approach that has been investigated in \cite{Bardeen:1981df,Liu:2016idz}.
As a further application, the techniques we developed in this work might be useful in the study of partial breaking of supersymmetry \cite{Bagger:1996wp,Ferrara:2014oka} and to understand its relation to dualities \cite{Broedel:2012gf}.

\section{The ${\cal N}=1$ goldstino in superspace} 
\label{sec:N=1_goldstino_in_superspace}

In this section we review the original formulation of the Volkov--Akulov model and we rephrase it in superspace, using the Samuel--Wess formalism.
An explicit proof of the equivalence of these two models is given directly in superspace.
The Rocek and the Komargodski--Seiberg constrained superfield models are then introduced and related to the previous ones.
This analysis has to be considered as a warm-up for the following sections, where we will promote our construction to extended supersymmetry.
In the calculations we adopt the conventions of \cite{Wess:1992cp}.

The Volkov--Akulov model \cite{Volkov:1973ix} is the prototype  of a spontaneously-broken minimal ${\cal N}=1$ theory with only a goldstino.
Given a goldstino field $\lambda^\alpha(x)$, its supersymmetry transformation is non-linearly realized and inhomogeneous, the constant term being proportional to the supersymmetry parameter $\epsilon^\alpha$ 
\begin{equation}
	\label{susygold}
	\delta\lambda^\alpha=f \epsilon^\alpha-\frac{i}{f}\left(\lambda \sigma^m\bar\epsilon-\epsilon\sigma^m\bar\lambda\right)\partial_m\lambda^\alpha .
\end{equation}
The goldstino can be used to define the invariant differential form
\begin{equation}
	\label{supervielb}
	dx^m {A_m}^a=dx^m\left[\delta_m^a-\frac{i}{f^2}\partial_m\lambda\sigma^a\bar\lambda+\frac{i}{f^2}\lambda\sigma^a\partial_m\bar\lambda\right]
\end{equation}
and to construct the Lagrangian
\begin{equation}
	{\cal L}=-f^2\det {A_m}^a=-f^2-i\left(\lambda\sigma^m\partial_m\bar\lambda-\partial_m\lambda\sigma^m\bar\lambda\right)+O(f^{-2}),
\end{equation}
where $f$ is the supersymmetry breaking scale with mass dimension 2.
This Lagrangian is invariant under supersymmetry, because 
\begin{equation}
	\label{LVA}
	\delta \det A = -\frac{i}{f}\partial_m\left[\left(\lambda \sigma^m\bar\epsilon-\epsilon\sigma^m\bar\lambda\right)\det A\right] .
\end{equation}

We are now going to rewrite the Volkov--Akulov model in superspace.
In our conventions the algebra of the $\mathcal{N}=1$ superspace derivatives is 
\begin{eqnarray}
	\{ D_\alpha , \bar D_{\dot \beta} \} = -2i\, \sigma^m_{\alpha \dot \beta} \partial_m  \ , \qquad \{ D_\alpha , D_\beta \} = 0 .
\end{eqnarray}
A superfield representation for the goldstino can be derived by considering a spinor superfield $\Lambda_\alpha$ satisfying the constraints 
\begin{eqnarray}
	\begin{split}
		\label{VASWrepr}
		D_\alpha \Lambda_\beta & = f\,\epsilon_{\beta\alpha}+\frac{i}{f}\,\sigma^m_{\alpha\dot\rho}\bar\Lambda^{\dot\rho}\partial_m\Lambda_\beta,\\
		\bar D_{\dot \alpha} \Lambda_\beta & = -\frac{i}{f}\,\Lambda^\rho \sigma^m_{\rho\dot\alpha}\partial_m\Lambda_\beta.
	\end{split}
\end{eqnarray}
This is a representation similar to the one introduced by Samuel--Wess 
in \cite{Samuel:1982uh}.
Actually the two are related by a field redefinition as we will see later.
More details on the role of these constraints in the theory of non-linear realizations can be found in \cite{Ivanov:1978mx,Ivanov:1982bpa}.
By construction, the goldstino is accomodated in the lowest component of $\Lambda_\alpha$ 
\begin{eqnarray}
	\Lambda_\alpha | = {\lambda_\alpha} 
\end{eqnarray}
and, due to the particular choice of the representation, its supersymmetry transformation is given precisely by \eqref{susygold}.
The superfield $\Lambda_\alpha$ can be used to build the superspace Lagrangian 
\begin{eqnarray}
	\label{LSW}
	{\cal L} = -\frac{1}{f^2} \int d^4 \theta \Lambda^2 \bar \Lambda^2 ,
\end{eqnarray} 
which reduces 
to the Volkov--Akulov action \eqref{LVA} at the component level.
In fact the equivalence of these two models can be proved directly in superspace, without working explicitly with the component fields.
Promote first the goldstino field $\lambda$ to a superfield $\Lambda$ satisfying \eqref{VASWrepr} 
and define the superspace analogous of the matrix $A_m^a$ to be
\begin{equation}
	\mathbb{A}_m^a = \delta_m^a-\frac{i}{f^2}\partial_m\Lambda\sigma^a\bar\Lambda+\frac{i}{f^2}\Lambda\sigma^a\partial_m\bar\Lambda .
\end{equation}
The superspace Lagrangian density has then the form 
\begin{equation}
	\label{LVAsuperspace}
	{\cal L}=-f^2\det {\mathbb{A}_m}^a |.
\end{equation} 
To prove the equivalence between \eqref{LSW} and \eqref{LVAsuperspace} notice that, due to the particular form of $\mathbb{A}_m^a$, up to boundary terms
\begin{equation}
	\int d^4\theta \Lambda^2\bar\Lambda^2 = \int d^4\theta \Lambda^2\bar\Lambda^2\det\mathbb{A}_m^a = \frac{1}{16} D^2 \bar D^2\left(\Lambda^2\bar\Lambda^2\det\mathbb{A}_m^a\right)|, 
\end{equation}
because terms in $\det\mathbb{A}_m^a$ containing either $\Lambda$ or $\bar\Lambda$ are annihilated by $\Lambda^2\bar\Lambda^2$ and only the constant term has an effective role in the computation.
Acting then with the covariant derivatives inside the parenthesis, the $\Lambda$ superfields are removed and, from the properties
\begin{equation}
	\begin{aligned}
		D_\rho \det\mathbb{A}_m^a             & = \frac{i}{f}\partial_m\left(\sigma^m_{\rho\dot\rho}\bar\Lambda^{\dot\rho}\det\mathbb{A}_m^a\right), \\
		\bar D^{\dot{\rho}}\det\mathbb{A}_m^a & =\frac{i}{f}\partial_m\left(\bar\sigma^{m\,\dot{\rho}\rho}\Lambda_\rho\det\mathbb{A}_m^a\right),     
	\end{aligned}
\end{equation}
the equivalence between the two Lagrangians follows up to total derivatives 
\begin{equation}
	\int d^4x d^4\theta\, \Lambda^2\bar\Lambda^2 = f^4 \int d^4x\det\mathbb{A}_m^a | \, .
\end{equation} 

{}From the Samuel--Wess superfield $\Lambda$ we can define a new chiral superfield
\begin{eqnarray}
	\Phi = - \frac{1}{4f^3} \bar D^2 \left( \Lambda^2 \bar \Lambda^2 \right).
\end{eqnarray}
In the representation \eqref{VASWrepr} it has the form  
\begin{eqnarray}
	\Phi = f^{-3}\,\Lambda^2 \left( f^2 - i \partial_m \Lambda \sigma^m \bar \Lambda  
	-  f^{-2}\bar \Lambda^2 \partial_m \Lambda \sigma^{mn} \partial_n \Lambda  \right).
\end{eqnarray}
This superfield $\Phi$ contains all the supersymmetry breaking information and satisfies the constraints 
\begin{eqnarray}
	\begin{aligned}
		\label{Rocekconstr}
		\Phi^2                  & = 0,        
		\\[0.1cm] 
		\Phi \bar D^2 \bar \Phi & = -4 f\Phi, 
	\end{aligned}
\end{eqnarray} 
which are exactly the constraints introduced in \cite{Rocek:1978nb}.
We can understand their role by assuming that $\Phi$ is an unconstrained chiral superfield.
By imposing the first constraint in \eqref{Rocekconstr}, the scalar component, namely the sgoldstino, is removed from the spectrum, while by imposing the second the supersymmetry breaking scale is fixed.
These constraints in fact reduce the number of independent component fields in the superfield 
and provide them as functions of the goldstino.

The goldstino inside the $\Phi$ superfield is defined as the component $D_\alpha \Phi |$ and, since
\begin{eqnarray}
	\label{DPhilambda}
	D_\alpha \Phi | = 2 \lambda_\alpha + \ldots\,, 
\end{eqnarray}
it is related to the field $\lambda_\alpha$ via a field redefinition.
For this reason the supersymmetric Lagrangian of the constrained $\Phi$ system,
\begin{eqnarray}
	{\cal L} = - f\int d^2 \theta \, \Phi ,
\end{eqnarray}
does not reduce directly to the Volkov--Akulov Lagrangian, as the Samuel--Wess Lagrangian does, 
because the goldstini have to be mapped into each other.
However, the proper field redefinition can be found by inverting the relation \eqref{DPhilambda} between the two goldstini.
Let us mention in passing that supersymmetry breaking from complex linear superfields  has been studied in \cite{Farakos:2013zsa,Farakos:2015vba} and the relation to the ${\cal N}=1$ goldstino superfields discussed here was also established.

At this point a comment on the role of Samuel--Wess representations in our discussion is in order.
We showed that the particular Samuel--Wess representation \eqref{VASWrepr} reproduces exactly the Volkov--Akulov model, either working in components or directly in superspace.
In fact there exists a second representation within the  Samuel--Wess formulation 
\begin{eqnarray}
	\label{SWchiralrepr}
	\begin{split}
		D_\alpha \Gamma_\beta & = 
		\,f\,\epsilon_{\beta\alpha},
		\\[0.1cm]
		\bar D_{\dot \alpha} \Gamma_\beta & = -  
		\,\frac{2i}{f}\,\Gamma^\rho \sigma^m_{\rho\dot\alpha}\partial_m\Gamma_\beta,
	\end{split}
\end{eqnarray}
which appeared in \cite{Samuel:1982uh}.
In superfield notation the supersymmetric Lagrangian is still the same as the one in \eqref{LSW}: 
\begin{equation}
	{\cal L} = -\frac{1}{f^2} \int d^4\theta \, \Gamma^2 \, \bar\Gamma^2 .
\end{equation}
To move from one representation to the other, set 
\begin{equation}
	\label{LambdaGamma}
	\Gamma_\alpha = - 2 f \, \frac{D_\alpha \bar D^2 ( \Lambda^2 \bar \Lambda^2 ) }{D^2 \bar D^2 ( \Lambda^2 \bar \Lambda^2 ) } 
\end{equation} 
and, using this relation, the equivalence between the two Lagrangians can be proved directly, because 
\begin{equation}
	\Gamma^2\bar\Gamma^2 = \Lambda^2 \bar \Lambda^2 .
\end{equation}

The approaches of  \cite{Samuel:1982uh} and \cite{Rocek:1978nb}, although being related by a field redefinition and thus being not exactly equivalent, have a common property: they describe a superfield containing only the goldstino and the supersymmetry breaking scale.
Fixing the supersymmetry breaking scale may be too restrictive for many purposes and therefore a less constrained approach has been developed.
To construct a superfield where the sgoldstino is removed from the spectrum, but the auxiliary field $F$ is still there, the second condition in \eqref{Rocekconstr} can be relaxed.
Indeed by imposing only
\begin{equation}
	X^2=0 \qquad \Longleftrightarrow \qquad X D_\alpha X=0,
\end{equation}
where the biconditional statement holds if $F$ acquires a non-vanishing vacuum-expectation-value, we have 
\begin{eqnarray}
	X = \frac{G^2}{2 F} + \sqrt 2 \theta G + \theta^2 F  \,,
\end{eqnarray} 
where $X$ is a chiral superfield with goldstino component $G_\alpha$.
This direction was followed in \cite{Casalbuoni:1988xh}, where the decoupling of the sgoldstino was explored.
In \cite{Komargodski:2009rz} the superfield $X$ was conjectured to be the IR limit of the superfield violating the supercurrent conservation equation in a supersymmetry-breaking setup\footnote{See \cite{Dudas:2011kt} for further discussions on the IR limit.}.
The minimal Lagrangian for $X$ is 
\begin{eqnarray}
	\label{LX1}
	{\cal L}  = \int d^4 \theta \, X \bar X + \left( f \int d^2 \theta \, X  + c.c. \right) .
\end{eqnarray}
The standard procedure to recover the Volkov--Akulov model from \eqref{LX1} consists in extracting the component fields, integrating out the auxiliary field $F$ and performing a field redefinition between the goldstini.
We present here a different procedure, relating the formalism of Komargodski--Seiberg to the one of Samuel--Wess directly in superspace.
Define the superfield
\begin{equation}
	\label{gamma}
	\Gamma_\alpha = -2\,f\,\frac{D_\alpha X}{D^2 X} \qquad \text{with} \qquad X^2=0.
\end{equation}
This is a Samuel--Wess superfield satisfying the representation \eqref{SWchiralrepr}.
Using \eqref{gamma}, the Ko\-mar\-godski--Seiberg Lagrangian \eqref{LX1} can be written as 
\begin{equation}
	{\cal L}  = \frac{1}{16 f^4} \int d^4 \theta \, \Gamma^2 \bar \Gamma^2 
	\left( D^2 X \bar D^2 \bar X  + 4 f D^2 X + 4 f \bar D^2 \bar X \right) .
\end{equation} 
Once the superspace integration is performed and up to boundary terms, this Lagrangian is equal to 
\begin{equation}
	\label{XNL33}
	{\cal L} = 
	\left(  \mathbb{F} \, \overline{ \mathbb{F}} +  f  \, \mathbb{F} +  f \, \overline{ \mathbb{F}} \right) 
	\det\mathbb{A}_m^a | ,
\end{equation} 
where we defined the superfield 
\begin{equation}
	\mathbb{F} = -\frac{1}{16 f^2}  (D - i \sigma^n \bar \Lambda \partial_n )^2 (\bar D + i \Lambda \sigma^n  \partial_n )^2 \left( X \, \bar \Gamma^2 \right).
\end{equation}
Integrating out $\mathbb{F}|$ gives
\begin{equation}
	\mathbb{F}|=-f 
\end{equation}
and, after substituting it back into the Lagrangian, \eqref{XNL33} reduces exactly to \eqref{LVAsuperspace}.
This concludes the equivalence between the models \eqref{LVAsuperspace} and \eqref{LX1}.
Notice that to relate the two formulations, the integration of a complex scalar is expected, because the Komargodski--Seiberg model contains also the auxiliary field $F$ in the Lagrangian.
More details on this proof can be found in  appendix \ref{appendixB}, where the origin of the scalar superfield $\mathbb{F}$ is explained.
We should mention here that a similar method was used in \cite{Luo:2009ib,Liu:2010sk} to prove the equivalence between various formulations and later in \cite{Kuzenko:2010ef} the same equivalence was studied in component form.

We conclude that all known realizations of the ${\cal N}=1$ goldstino model are equivalent and we proved it in superspace, therefore making the procedure more transparent.

\section{The ${\cal N}=2$ goldstini in superspace} 
\label{sec:N=2_goldstini_in_superspace}

In this section we promote the Samuel--Wess formalism to ${\cal N}=2$ superspace and we use it to find the appropriate constrained superfield approach\footnote{For a different formulation of the theory of ${\cal N}=2$ goldstini, see \cite{Kuzenko:2011ya}.}.
The first part of the analysis is similar to \cite{Kandelakis:1986bj}, but it is worked out in a different representation.
Once the minimal set of constraints is determined, the theory is reformulated in ${\cal N}=1$ language and the complete expression for the Lagrangian in the $X,Y$ system is given for the first time.
Some of the demonstrations of this section are omitted, because we are going to prove the analogous results directly for general ${\cal N}$.

\subsection{$\mathcal{N}=2$ superspace} 
\label{sub:N=2_breaking_in_N=2_language}

The algebra satisfied by the $\mathcal{N}=2$  superspace derivatives without central charges is
\begin{eqnarray}
	\begin{split}
		\{ D_\alpha , \bar D_{\dot \alpha} \} &=\{ \tilde D_\alpha , \bar{\tilde D}_{\dot \alpha} \} = -2i\, \sigma^m_{\alpha \dot \alpha} \partial_m , 
		\\[0.2cm]
		\{ D_\alpha , D_\beta \} &= \{ \tilde D_\alpha , \tilde D_\beta \}= \{ D_\alpha , \tilde D_\beta \} = \{ D_\alpha , \bar{\tilde D}_{\dot \beta} \}=  0 ,
	\end{split}
\end{eqnarray}
where $\tilde D_\alpha$ generates the second supersymmetry.

The first step consists in determining the (minimal) set of constraints needed to remove from the spectrum all the undesired component fields.
To this purpose we exploit the Samuel--Wess formalism, following the procedure outlined in the previous section for the $\mathcal{N}=1$ theory.
When supersymmetry is completely broken, we need two goldstini and therefore we define two spinor superfield $\Lambda_\alpha$ and $\tilde \Lambda_\alpha$ satisfying the constraints 
\begin{eqnarray}
	\begin{split} 
		D_\alpha \Lambda_\beta & = f\,\epsilon_{\beta\alpha} +\frac{i}{f}\sigma^m_{\alpha\dot{\rho}}\bar{\Lambda}^{\dot\rho}\partial_m\Lambda_\beta \, ,  
		\qquad 
		\bar D_{\dot \alpha} \Lambda_\beta  = -\frac{i}{f}\, \sigma^m_{\rho \dot \alpha} \Lambda^{\rho} \partial_m \Lambda_\beta ,
		\\[0.1cm] 
		\tilde D_\alpha \tilde \Lambda_\beta & =f\, \epsilon_{\beta\alpha} +\frac{i}{f}\sigma^m_{\alpha\dot{\rho}}\bar{\tilde\Lambda}^{\dot\rho}\partial_m\tilde{\Lambda}_\beta \, ,
		\qquad 
		\bar{\tilde D}_{\dot \alpha} \tilde \Lambda_\beta  = -\frac{i}{f}\, \sigma^m_{\rho \dot \alpha} \tilde \Lambda^{\rho} \partial_m \tilde \Lambda_\beta 
	\end{split}
\end{eqnarray}
and 
\begin{eqnarray}
	\begin{split} 
		\tilde D_\alpha \Lambda_\beta & = \frac{i}{f}\sigma^m_{\alpha\dot{\rho}}\bar{\tilde{\Lambda}}^{\dot\rho}\partial_m\Lambda_\beta \, ,\phantom{-}
		\qquad 
		D_\alpha \tilde \Lambda_\beta  = \frac{i}{f}\sigma^m_{\alpha\dot{\rho}}\bar{\Lambda}^{\dot\rho}\partial_m\tilde{\Lambda}_\beta ,
		\\[0.1cm] 
		\bar{\tilde D}_{\dot \alpha} \Lambda_\beta & = -\frac{i}{f}\, \sigma^m_{\rho \dot \alpha} \tilde \Lambda^{\rho} \partial_m \Lambda_\beta \, , 
		\qquad 
		\bar D_{\dot \alpha} \tilde \Lambda_\beta  = -\frac{i}{f}\, \sigma^m_{\rho \dot \alpha} \Lambda^{\rho} \partial_m \tilde \Lambda_\beta .
	\end{split}
\end{eqnarray}
The superfields $\Lambda_\alpha$ and $\tilde \Lambda_\alpha$ are the Samuel--Wess goldstino superfields for the broken supersymmetries.
The only independent component fields they have are the goldstini, defined as 
\begin{equation} 
	\Lambda_\alpha |_{\theta=\tilde \theta=0} =\lambda_\alpha \, , \qquad 
	\tilde \Lambda_\alpha |_{\theta=\tilde \theta=0} =\tilde{\lambda}_\alpha  .
\end{equation}
Their supersymmetry transformations are non-linearly realized 
\begin{eqnarray}
	\begin{split}
		\delta \lambda_\alpha  =& f\, \epsilon_\alpha-\frac{i}{f}\left(\lambda \sigma^m\bar\epsilon-\epsilon\sigma^m\bar\lambda\right)\partial_m\lambda_\alpha-\frac{i}{f}\left(\tilde\lambda \sigma^m\bar{\tilde\epsilon}-\tilde\epsilon\sigma^m\bar{\tilde\lambda}\right)\partial_m{\lambda}_\alpha,
		\\[0.1cm]
		\delta \tilde \lambda_\alpha  =&  f\,\tilde \epsilon_\alpha 
		-\frac{i}{f}\left(\tilde\lambda \sigma^m\bar{\tilde\epsilon}-\tilde\epsilon\sigma^m\bar{\tilde\lambda}\right)\partial_m \tilde \lambda_\alpha 
		-\frac{i}{f}\left(\lambda \sigma^m\bar\epsilon-\epsilon\sigma^m\bar\lambda\right)\partial_m \tilde \lambda_\alpha ,
	\end{split}
\end{eqnarray} 
where $\epsilon_\alpha$, $\tilde \epsilon_\alpha$ are the $\mathcal{N}=2$ supersymmetry parameters.
As discussed more carefully in the general ${\cal N}$ section, the supersymmetric and $\textup{U}(2)_R$-invariant Lagrangian can be written as
\begin{equation}
	{\cal L} =  -\frac{1}{f^6}\int d^4 \theta d^4\tilde\theta\, \Lambda^2\tilde{\Lambda}^2\bar\Lambda^2\bar{\tilde{\Lambda}}^2
\end{equation}
and, after projecting to components, we find  
\begin{eqnarray}
	{\cal L} = -f^2-i(\lambda\sigma^m\partial_m\bar\lambda-\partial_m\lambda \sigma^m\bar\lambda)-i(\tilde{\lambda}\sigma^m\partial_m\bar{\tilde\lambda}-\partial_m\tilde{\lambda} \sigma^m\bar{\tilde\lambda})+ O(f^{-2}) .
\end{eqnarray} 

We develop now the constrained superfield approach in ${\cal N}=2$ superspace.
Define first a chiral superfield $\Phi$
\begin{eqnarray} 
	\Phi = \frac{1}{16f^7} 
	\bar D^{2} \bar{ \tilde D}^{2}  
	\left( \Lambda^{2}  \bar \Lambda^{2} \tilde \Lambda^{2}  \bar{ \tilde \Lambda}^{2} \right) , 
\end{eqnarray}
containing the supersymmetry breaking information.
We are going to use it to derive the appropriate set of constraints to remove all the undesired component fields.
This superfield can be expressed in the form
\begin{eqnarray}
	\Phi = f^{-5}\Lambda^2 \tilde \Lambda^2 \left( f^2 - i \partial_a \Lambda \sigma^a \bar \Lambda 
	- i \partial_a \tilde \Lambda \sigma^a \bar{ \tilde \Lambda}   + O(f^{-2}) \right) 
\end{eqnarray}
and by construction one can verify that
\begin{eqnarray}
	\bar D_{\dot \alpha} \Phi = \bar{\tilde D}_{\dot \alpha} \Phi =0 .
\end{eqnarray} 
The goldstini can be found in the component fields 
\begin{eqnarray}
	\begin{split}
		-\frac14 D^2 \tilde D_\alpha \Phi |_{\theta=\tilde \theta=0} =& 2 \tilde \lambda_\alpha + \ldots ,
		\\
		-\frac14 \tilde  D^2  D_\alpha \Phi |_{\theta=\tilde \theta=0} =& 2 \lambda_\alpha  + \ldots ,
	\end{split}
\end{eqnarray}
where dots stand for terms with more fermions.
By direct inspection we find that $\Phi$ satisfies the following set of constraints 
\begin{eqnarray}
	\label{phi2}
	\begin{split}
		\Phi^2 &= 0,
		\\[0.09cm]
		\Phi D_\alpha  \Phi &= \Phi \tilde D_\alpha  \Phi   = 0 ,
		\\[0.1cm]
		\Phi \tilde D_\alpha D_\beta \Phi & 
		= \Phi \tilde D_\alpha \tilde D_\beta \Phi 
		= \Phi D_\alpha D_\beta \Phi = 0 ,
		\\[0.11cm]
		\Phi \tilde D_\alpha D_\beta \tilde D_\gamma \Phi & 
		= \Phi \tilde D_\alpha  D_\beta D_\gamma \Phi=0,
	\end{split}
\end{eqnarray}
together with the property 
\begin{eqnarray}
	\label{phivev}
	\bar \Phi D^2 \tilde D^2 \Phi = 16\,f \, \bar \Phi ,
\end{eqnarray}
implying again the fact that the highest component of the superfield contains the supersymmetry-breaking scale.
The $\mathcal{N}=2$ goldstini Lagrangian for $\Phi$ can be written as 
\begin{eqnarray}
	\label{lagrangian2phi} 
	{\cal L} = -f \int d^2 \theta d^2 \tilde \theta \, \Phi .
\end{eqnarray}

By imposing the constraints \eqref{phi2} to a generic ${\cal N}=2$ chiral superfield, we remove all of its components except the goldstini and the auxiliary field containing the supersymmetry breaking scale.
The important point is that only the last constraint in \eqref{phi2} is really essential.
We demonstrate this statement in the general ${\cal N}$ section, solving the constraint explicitly in superspace and verifying that the unique solution contains only ${\cal N}$ goldstini and the correct auxiliary field.
Consider therefore a ${\cal N}=2$ chiral superfield ${\cal X}$ 
\begin{eqnarray}
	\bar D_{\dot \alpha} {\cal X}  = \bar{\tilde D}_{\dot \alpha} {\cal X}=0 .
\end{eqnarray} 
Imposing the constraint
\begin{equation}
	{\cal X} \tilde D_\alpha D_\beta \tilde D_\gamma {\cal X}
	= {\cal X} \tilde D_\alpha  D_\beta D_\gamma {\cal X}=0 
\end{equation}
and solving it in superspace we obtain the unique solution
\begin{equation}
	\label{XN2sol}
	{\cal X}=\frac{1}{4}\frac{G^2\,\tilde{G}^2}{{\cal F}^3}.
\end{equation}
In this formula the ${\cal N}=2$ superfield ${\cal F}$ is defined such that
\begin{eqnarray}
	\label{FN2}
	F={\cal F}|_{\theta=\tilde \theta=0} = \frac{1}{16} D^2 \tilde D^2 {\cal X} |_{\theta=\tilde \theta=0}
\end{eqnarray}
is the complex scalar auxiliary field, while the superfields $G_{\alpha},\tilde{G}_\alpha$ are defined such that
\begin{eqnarray}
	\label{GN2}
	\begin{split}
		g_\alpha=\tilde G_\alpha|_{\theta=\tilde \theta=0} =&  -\frac{1}{4 \sqrt 2} D^2 \tilde D_\alpha {\cal X} |_{\theta=\tilde \theta=0},
		\\
		{\tilde{g}}_\alpha=G_\alpha |_{\theta=\tilde \theta=0} =&  - \frac{1}{4 \sqrt 2} \tilde D^2 D_\alpha {\cal X} |_{\theta=\tilde \theta=0} ,
	\end{split}
\end{eqnarray}
are the goldstini.
From the explicit form of the solution \eqref{XN2sol} and from the properties \eqref{FN2}, \eqref{GN2} one can see that all component fields in ${\cal X}$
are effectively solved in terms of the auxiliary scalar ${F}$ and of the two goldstini $g_\alpha$, $\tilde g_\alpha$.

To have a more direct understanding of what is going on, we can introduce chiral coordinates
\begin{eqnarray}
	y^m = x^m  + i \theta \sigma^m \bar \theta 
	+ i \tilde \theta \sigma^m \bar{ \tilde \theta}
\end{eqnarray} 
and expand ${\cal X}$ as 
\begin{equation}
	\begin{aligned}
		{\cal X} & = \frac{g^2\,\tilde{g}^2}{4F^3}+\frac{\tilde{g}^2\, g^\alpha}{\sqrt{2}F^2}\theta_\alpha+\frac{\tilde{g}^2}{2F}\theta^2 + \frac{{g}^2\, \tilde{g}^\alpha}{\sqrt{2}F^2}\tilde{\theta}_\alpha + 2\,\frac{\tilde{g}^\alpha\tilde{\theta}_\alpha\,g^\beta\theta_\beta }{F}+\frac{g^2}{2F}\tilde{\theta}^2+ \\[0.1cm]
		        & +\sqrt{2}\,g^\alpha\tilde{\theta}_\alpha \theta^2+ \sqrt{2}\,g^\alpha\theta_\alpha \tilde{\theta}^2+F\,\theta^2\tilde{\theta}^2.
	\end{aligned}
\end{equation}
The Lagrangian for the supersymmetry breaking sector is now
\begin{eqnarray}
	\label{lagrangian2X}
	{\cal L} = \int d^4 \theta d^4 \tilde \theta \, {\cal X} \bar {\cal X} 
	+ \left( f \int d^2 \theta d^2 \tilde \theta \, {\cal X}  + c.c. \right) 
\end{eqnarray}
and to get the explicit goldstini action, after projecting to components, one has to solve the equation of motion for the auxiliary field $F$ via an iterative procedure giving 
\begin{eqnarray}
	\label{X2F}
	{ F} = - f + \ldots 
\end{eqnarray}
and then replace the solution back into \eqref{lagrangian2X}.

\subsection{$\mathcal{N}=1$ superspace}

We are now in the process of formulating the previous results in $\mathcal{N}=1$ language, which is the most useful one for practical applications.
In doing so, we will show that we have various ways of describing the interactions of the goldstini, depending on the realization of the second goldstino as the fermion of a chiral or tensor multiplet, or as the upper component of a vector multiplet.
This could be useful if one wants to understand the (possible) ultraviolet completions in terms of hyper, vector or tensor multiplets.
Expand first ${\cal X}$ as 
\begin{eqnarray}
	\label{XinNis1}
	{\cal X} = S(y,\theta) + \sqrt 2 \,  \tilde \theta^\beta W_\beta (y,\theta) + \tilde \theta^2 X(y,\theta)  ,
\end{eqnarray} 
where $S$, $W_\alpha$ and $X$ are $\mathcal{N}=1$ chiral superfields
\begin{eqnarray}
	\bar D_{\dot \alpha} S = 0 , \ \bar D_{\dot \alpha} W_\alpha = 0 , \ \bar D_{\dot \alpha} X = 0.
\end{eqnarray}
The first supersymmetry acts on these superfields as usual,
\begin{eqnarray}
	\delta_1 {\cal O} = \epsilon^\alpha D_\alpha {\cal O}  + \bar \epsilon_{\dot \alpha} \bar D^{\dot \alpha} {\cal O} ,    
\end{eqnarray}
and one can derive from here the supersymmetry transformations of the component fields.
The second supersymmetry acts by transforming the $\mathcal{N}=1$ superfields into each other 
\begin{eqnarray}
	\label{secondsusy}
	\begin{split}
		\delta_2 S &=\sqrt 2 \tilde \epsilon^\alpha W_\alpha,
		\\[0.1cm]
		\delta_2 W_\alpha&= \sqrt 2 i \sigma_{\alpha \dot \alpha}^m \bar{\tilde \epsilon}^{\dot \alpha} \partial_m S + \sqrt 2 \tilde \epsilon_\alpha X ,
		\\[0.1cm]
		\delta_2 X &= \sqrt 2 i \bar{\tilde \epsilon}_{\dot \alpha} \bar \sigma^{m \dot \alpha \alpha} \partial_m W_\alpha .
	\end{split}
\end{eqnarray} 
In particular the auxiliary field acquiring a non-vanishing vacuum-expectation-value is now expressed as 
\begin{eqnarray}
	{ F}  = -\frac14  D^2 X |   
\end{eqnarray}
and, from the supersymmetry transformations
\begin{eqnarray}
	\begin{split}
		\delta_1 g _\alpha  &= \frac{1}{\sqrt 2} \delta_1 D_\alpha X | = \sqrt 2 \epsilon_\alpha { F} + \ldots  ,  
		\\
		\delta_2 \tilde g _\alpha  &= -\frac{1}{4} \delta_2 D^2 W_\alpha | = \sqrt 2 \tilde \epsilon_\alpha { F} + \ldots ,  
	\end{split}
\end{eqnarray}
one can understand that the goldstini are accomodated inside the superfields $X$ and $W_\alpha$.

Inserting the explicit expression \eqref{XinNis1} in \eqref{phi2}, we find a large number of constraints for the ${\cal N}=1$ superfields 
\begin{eqnarray}
	\label{constr1}
	\begin{split}
		S^2=W^2=X^2 &= 0 ,
		\\
		S D_\beta X&= 0 ,
		\\
		W_\alpha D_\beta X  &=0, 
		\\
		W^\alpha D^2 W_\alpha &= 2S D^2 X  .
	\end{split}
\end{eqnarray} 
As we have already argued however the minimal number of constraints has to be very small, just one for the superfield ${\cal X}$ living in the full ${\cal N}=2$ superspace, and in fact only the following ${\cal N}=1$ constraints can be thought as fundamental 
\begin{eqnarray}
	\label{fondconstrN2inN1}
	S =\frac{X}{2} \frac{ (D^2 W)^2}{(D^2 X)^2} ,\qquad W_\alpha = X \frac{D^2 W_\alpha }{D^2 X}, \qquad   X^2=0.
\end{eqnarray}
From the first of these constraints, in particular, it is manifest that  $S$ is entirely removed and expressed in terms of $W_\alpha$ and $X$, containing in turn the goldstini and the auxiliary field ${F}$.

We can now write the Lagrangian \eqref{lagrangian2X} in the ${\cal N}=1$ constrained superfield language, replacing $S$ with its expression in terms of $W_\alpha$ and $X$.
The result is the low energy theory of an $\mathcal{N}=2$ spontaneously broken supersymmetry generalizing the Volkov--Akulov model 
\begin{eqnarray}
	\label{N2LL}
	{\cal L} = 
	\int d^4 \theta \left( X \bar X 
	- \Big{|}\partial_m \left( \! \frac{X}{2} \frac{ (D^2 W)^2}{(D^2 X)^2} \right) \Big{|}^2
	+ i \partial_m W^\alpha \sigma^m_{\alpha \dot \alpha} \bar W^{\dot \alpha}  \right) 
	+ f \left( \int d^2 \theta X + c.c. \right) .
\end{eqnarray}
Notice that, on top of the manifest $\mathcal{N}=1$ supersymmetry, \eqref{N2LL} has a second supersymmetry given by 
\begin{eqnarray}
	\label{secondsusynilpotent}
	\begin{split}
		\delta_2 W_\alpha&= \sqrt 2 i \sigma_{\alpha \dot \alpha}^m \bar{\tilde \epsilon}^{\dot \alpha} \partial_m \left( \! \frac{X}{2} \frac{ (D^2 W)^2}{(D^2 X)^2} \right) 
		+ \sqrt 2 \tilde \epsilon_\alpha X ,
		\\
		\delta_2 X &= \sqrt  2  i \bar{\tilde \epsilon}_{\dot \alpha} \bar \sigma^{m \dot \alpha \alpha} \partial_m W_\alpha ,
	\end{split}
\end{eqnarray} 
and since the superfields $W_\alpha$ and  $X$ are constrained, this second supersymmetry is non-linearly realized.
In component form the Lagrangian \eqref{N2LL} reduces to 
\begin{eqnarray}
	\label{N2LLL}
	{\cal L} = {F} \bar {F} + f {F} + f \bar { F}  
	+ i \partial_m g^\alpha \sigma^m_{\alpha \dot \alpha} \bar g^{\dot \alpha}
	+ i \partial_m \tilde g^\alpha \sigma^m_{\alpha \dot \alpha} \bar{ \tilde g}^{\dot \alpha} 
	+ \text{higher order fermion terms}.
\end{eqnarray}
As a final step one can integrate out $F$ and replace its expression into \eqref{N2LLL}, obtaining a theory for the $\mathcal{N}=2$ supersymmetry breaking with only goldstini.

An alternative way to describe this theory is by defining the spinor superfield 
\begin{equation}
	H_{\dot \alpha} = \frac{\bar D^2 \bar W_{\dot \alpha}}{\bar D^2 \bar X} .
\end{equation} 
This is a chiral superfield,
\begin{equation}
	\bar D_{\dot \beta} H_{\dot \alpha}  = 0 ,
\end{equation}
satisfying the property
\begin{equation}
	\label{XH}
	\bar D_{\dot \beta} \left( X \bar H_\alpha  \right) = 0  .
\end{equation}
Since $H_{\dot \alpha}$ is chiral, it is known \cite{Komargodski:2009rz} that the constraint \eqref{XH} removes its higher components leaving the lowest one, namely the goldstino of the second supersymmetry, unconstrained.
The low energy theory of an $\mathcal{N}=2$ spontaneously broken supersymmetry can be expressed in this $X,\,H_{\dot{\alpha}}$ system as
\begin{eqnarray}
	\label{N2LL33}
	{\cal L} = 
	\int d^4 \theta \left( X \bar X 
	- \Big{|}\partial_m \left( \! \frac{X \bar H^2}{2}  \right) \Big{|}^2
	+ i \partial_m (X \bar H^\alpha) \sigma^m_{\alpha \dot \alpha} (\bar X H^{\dot \alpha} ) \right) 
	+ f \left( \int d^2 \theta X + c.c. \right) .
\end{eqnarray} 

In fact there exists another way to describe this model.
The Lagrangian \eqref{N2LL} describes a theory with two chiral $\mathcal{N}=1$ constrained superfields where all the components have been removed except the goldstini and the auxiliary field breaking supersymmetry.
This type of theories have been analyzed in terms of orthogonal nilpotent superfields
\begin{equation}
	X^2=XY=0, \qquad \text{$X,Y$ chiral} \qquad (Y^3=0).
\end{equation}
Since all these theories have the same physical content, however, there must be a way to rewrite \eqref{N2LLL} in terms of the $\mathcal{N}=1$ constrained superfield $X$ and $Y$.
In the presence of an $\mathcal{N}=1$ supersymmetry-breaking constrained chiral superfield $X$ satisfying
\begin{eqnarray}
	X^2 = 0  \qquad \Longleftrightarrow \qquad X D_\alpha X=0
\end{eqnarray}
we define therefore a chiral superfield $Y$,
\begin{eqnarray}
	\bar D_{\dot \alpha} Y = 0 ,
\end{eqnarray}
and we remove its scalar and auxiliary field components imposing the constraints
\begin{eqnarray}
	\label{newY}
	\begin{split}
		XY &= 0,
		\\
		\bar X D^2 Y  &= 0 .
	\end{split}
\end{eqnarray}
These constraints can be solved to obtain
\begin{eqnarray}
	\label{ConstrY}
	\begin{split}
		Y & = -2 \frac{D^\alpha X D_\alpha Y}{D^2 X} - X \frac{D^2 Y}{D^2 X},
		\\[0.1cm]
		D^2 Y&= \frac{- 16 \bar X \partial^2 Y + 8 i \bar D_{\dot \alpha} \bar X \partial^{\alpha \dot \alpha} D_\alpha Y}{\bar D^2 \bar X}   , 
	\end{split}
\end{eqnarray}
where the notation $\partial_{\alpha \dot \alpha} \equiv \sigma^m_{\alpha \dot \alpha} \partial_m$ has been used.
From \eqref{ConstrY} it can be understood that the lowest component of the $Y$ superfield, namely $Y|$, and the auxiliary field $D^2 Y|$ are removed from the spectrum and expressed in terms of the fermion $D_\alpha Y|$.
The chiral superfield $Y$ contains therefore only one fermion.
Since also the superfield $W_\alpha$ contains only one fermion, it should be possible to use it to build a superfield having exactly the properties of $Y$. The expression 
\begin{eqnarray}
	\label{WY}
	Y = - \frac{1}{\sqrt 2} D^\alpha W_\alpha + \sqrt 2 \frac{\bar D^{\dot \rho} \bar X \bar D_{\dot \rho} D^\rho W_\rho}{\bar D^2 \bar X} ,
\end{eqnarray}
satisfying \eqref{newY} when $W_\alpha$ satisfies the second constraint in \eqref{fondconstrN2inN1}, is the desired one.

To rewrite the Lagrangian \eqref{N2LL} in the $X,Y$ system we have first to invert \eqref{WY}, in order to express $W_\alpha$ in terms of the chiral constrained superfield $Y$ 
\begin{eqnarray}
	\label{WofY}
	\begin{split}
		W_\beta = & 2\sqrt 2 \frac{X D_\beta Y}{D^2 X} 
		+ 16 \sqrt 2 i X \frac{D^\rho Y}{D^2 X} \frac{\bar D^{\dot \rho} \bar X}{\bar D^2 \bar X} \partial_{\rho \dot \rho} \left( \frac{D_\beta X}{D^2 X} \right) -
		\\
		& - 128 \sqrt 2 X \frac{D^\sigma Y}{D^2 X} \frac{\bar D^{\dot \sigma} \bar X}{\bar D^2 \bar X} \partial_{\sigma \dot \sigma} \left( \frac{D^\rho X}{D^2 X} \right) 
		\frac{\bar D^{\dot \rho} \bar X}{\bar D^2 \bar X} \partial_{\rho \dot \rho} \left( \frac{D_\beta X}{D^2 X} \right) .
	\end{split}
\end{eqnarray} 
We can then replace \eqref{WofY} in \eqref{N2LL} to obtain the Lagrangian
\begin{eqnarray}
	{\cal L}  = \int d^4 \theta \left( X \bar X + Y \bar Y [ 1 + {\cal A} ] + S \partial^2 \bar S \right) 
	+ f \left( \int d^2 \theta X + c.c. \right) ,
\end{eqnarray}
where 
\begin{equation}
	\label{Aterm}
	{\cal A } = - 64 \frac{ \bar D_{\dot \gamma} \bar X \partial_{\rho}^{\dot \gamma} \bar D^{\dot \rho}  \bar X
		D_\gamma X \partial_{\dot \rho}^{\gamma}  D^\rho X }{|D^2 X|^4} 
\end{equation}
and 
\begin{equation}
	S = Y^2  \frac{ D^2 X}{ 
		\left( \delta_{\epsilon}^{\rho} 
		+ 8 i \frac{\bar D_{\dot \rho} \bar X \partial_{\epsilon}^{\dot \rho} D^{\rho} X}{|D^2 X|^2} \right) 
		\left( \delta^{\epsilon}_{\rho} 
		- 8 i \frac{\bar D_{\dot \gamma} \bar X \partial^{\dot \gamma \epsilon} D_\rho X}{|D^2 X|^2} \right) 
	} .
\end{equation}
This is the complete expression of the ${\cal N}=2$ supersymmetry breaking Lagrangian in the language of orthogonal nilpotent superfields.

\section{${\cal N}$ goldstini} 
\label{sec_N_goldstini_in_superspace}

In this section we generalize our results to an arbitrary number ${\cal N}$ of supersymmetries.
The logical thread is the same as in the previous section, but the steps are justified with more care.
General expressions for the component fields and for the Lagrangian are given and a way to organize the high number of removed fields is depicted.
Some technical and rather long calculations are reported in detail in the appendices.

\subsection{${\cal N}$ superspace}

The  algebra satisfied by the ${\cal N}$ superspace derivatives without central charges is
\begin{eqnarray}
	\begin{split}
		\{ D^I_\alpha , \bar D_{J\,\dot \alpha} \} &= -2i \, \delta^I_J \, \sigma^m_{\alpha \dot \alpha} \partial_m , 
		\\[0.1cm]
		\{ D^I_\alpha , D^J_\beta \} &= 0 , 
	\end{split}
\end{eqnarray}
where the indices $I,J$ run from 1 to $\mathcal{N}$ labelling the supersymmetries.
In particular lower indices refer to the fundamental of $\textup{U}({\cal N})_R$, while upper indices refer to the antifundamental.
Since there are now ${\cal N}$ broken supersymmetries, the theory contains ${\cal N}$ goldstini and therefore we define ${\cal N}$ spinor superfields $\Lambda_{I\,\alpha}$ satisfying the constraints
\begin{eqnarray}
	\label{NSWrepr}
	\begin{split} 
		D_\alpha^I \Lambda_{J\,\beta} & = f\,\epsilon_{\beta\alpha} \, \delta^I_J+\frac{i}{f}\sigma^m_{\alpha\dot{\rho}}\bar\Lambda^{I\,\dot{\rho}}\partial_m \Lambda_{J\,\beta} ,
		\\
		\bar D_{I\,\dot \alpha} \Lambda_{J\,\beta} & = -\frac{i}{f} \sigma^m_{\rho \dot \alpha} \Lambda^{\rho}_I
		\partial_m \Lambda_{J\,\beta} .
	\end{split}
\end{eqnarray}
The superfields $\Lambda_{I\,\alpha}$ are the Samuel--Wess goldstino superfields for the broken supersymmetries.
The only independent component fields are the goldstini, defined as 
\begin{eqnarray}
	\Lambda_{I\,\alpha} |_{\theta^I=0} = \lambda_{I\, \alpha}.
\end{eqnarray}
Their supersymmetry transformations are
\begin{eqnarray}
	\delta \lambda_{I\,\alpha} = f\epsilon_{I\,\alpha} - \frac{i}{f} \sum_J\left( \lambda_J \sigma^m \, \bar \epsilon^J-\epsilon_J\sigma^m\bar\lambda^J \right) \partial_m \lambda_{I\,\alpha}  ,
\end{eqnarray}  
where $\epsilon_{I\,\alpha}$ are the ${\cal N}$ supersymmetry parameters.
The supersymmetric and $\textup{U}({\cal N})_R$-invariant Lagrangian can be written in several equivalent ways
\begin{equation}
	\label{LN}
	\begin{aligned}
		\mathcal{L} & =-\frac{C^2_\mathcal{N}}{f^{4\mathcal{N} -2}}\int d^{4\mathcal{N}}\theta \det(\Lambda_{I_1}\Lambda_{I_2}\ldots \Lambda_{I_\mathcal{N}}) \det (\bar\Lambda^{J_1}\bar\Lambda^{J_2}\ldots\bar\Lambda^{J_\mathcal{N}})                                                                                                                
		\\[0.1cm]
		            & = - \frac{1}{f^{4\mathcal{N} -2}}\int d^{4\mathcal{N}}\! \theta \,  (\Lambda^{2}_1 \Lambda^{2}_2 \ldots\Lambda^{2}_\mathcal{N} )(\bar \Lambda^{2}_1\bar\Lambda^{2}_2\ldots\bar\Lambda^{2}_\mathcal{N})\equiv- \frac{1}{f^{4\mathcal{N} -2}}\int d^{4\mathcal{N}}\! \theta \,  \Lambda^{2\mathcal{N}} \bar \Lambda^{2 \mathcal{N}} 
		\\[0.1cm]
		            & \propto -\frac{1}{f^{4\mathcal{N} -2}}\int d^{4\mathcal{N}}\theta\! \left(\Lambda_{I\,\alpha}\Lambda^\alpha_J\bar\Lambda^I_{\dot{\alpha}}\bar{\Lambda}^{J\,\dot\alpha}\right)^\mathcal{N}                                                                                                                                         
	\end{aligned}
\end{equation}
where $C_\mathcal{N}$ is a normalization chosen in such a way that the first line in \eqref{LN} reduces to the second one.
The power of $\mathcal{N}$ in the last line is fixed by the requirement of having a minimal effective theory with only goldstini and whose component expansion starts with a constant term, in order to recover the positive constant breaking supersymmetry.
Indeed, the Lagrangian \eqref{LN} includes kinetic terms, the vacuum energy and higher order corrections essential for the non-linear realization.
For simplicity in what follows we use directly the more compact expression 
\begin{equation}
	{\cal L}=-\frac{1}{f^{4\mathcal{N}-2}}\int d^{4\mathcal{N}}\! \theta \,  \Lambda^{2\mathcal{N}} \bar \Lambda^{2 \mathcal{N}} 
\end{equation}
and, after projecting to components, we find
\begin{eqnarray}
	{\cal L} = -f^2 - \sum_I i (\lambda_I \sigma^m \partial_m\bar\lambda^I-\partial_m\lambda_I\sigma^m\bar\lambda^I) + O(f^{-2}).
\end{eqnarray} 

We develop now the constrained superfield approach in ${\cal N}$ superspace.
Define first the chiral superfield $\Phi$
\begin{eqnarray} 
	\label{Finn}
	\Phi = \frac{1}{(-4)^{\mathcal{N}}f^{4\mathcal{N}-1}} \bar D^{2\mathcal{N}} \left( \Lambda^{2\mathcal{N}} \bar \Lambda^{2 \mathcal{N}} \right) 
\end{eqnarray}
containing the supersymmetry-breaking information.
We are going to use it to derive the appropriate set of constraints to remove all the undesired component fields\footnote{We use the notation $(\Psi_{I})^{2{\cal N}}\equiv\Psi^{2{\cal N}}=\Psi_1^2\,\Psi_2^2\ldots \Psi_{{\cal N}}^2$ to indicate the product of ${\cal N}$ squared spinor superfields. The dummy index $I$ is not summed and, to avoid confusion, whether sums occur they are going to be explicitly written. In some formulas, the notation $\hat{I}$ is actually used to stress the fact that the index is fixed.}.

This superfield can be expressed in the form 
\begin{eqnarray} 
	\label{Phistruct}
	\Phi = f^{-(2\mathcal{N}+1)}\Lambda^{2\mathcal{N}} \left( f^2 + \sum_I\bar\Lambda^I(\ldots)+\sum_{I,J}\bar\Lambda^{I}\bar\Lambda^{J}(\ldots)+\ldots \right) ,
\end{eqnarray} 
where $(\ldots)$ contains terms with derivatives of the $\Lambda$ superfields, and by construction one can verify that
\begin{equation}
	\bar D_{I\,\dot{\alpha}}\Phi =0.
\end{equation}
The goldstini can be found in the component fields
\begin{equation}
	\frac{1}{(-4)^{{\cal N}-1}}(D^{I\neq J})^{2{\cal N}-2}D^J_{\alpha}\Phi|_{\theta^I=0}=2\lambda_{J \alpha}+\ldots,
\end{equation}
where dots stand for terms with more fermions.
By direct inspection of formula \eqref{Phistruct} we find that $\Phi$ satisfies the following set of constraints 
\begin{eqnarray}
	\label{phi2N}
	\begin{split}
		\Phi^2 &= 0,
		\\
		\Phi D^I_\alpha  \Phi &= 0, 
		\\
		\Phi D^I_\alpha D^J_\beta \Phi & =0 ,
		\\
		\Phi D^I_\alpha D^J_\beta D^K_\gamma \Phi & =0 ,
		\\
		\cdots & 
		\\
		\Phi (D^{I\neq J})^{2\mathcal{N}-2}D^J_\alpha  \Phi & =0 ,
	\end{split}
\end{eqnarray}
together with the property
\begin{eqnarray}
	\bar \Phi  (D^I)^{2\mathcal{N}} \Phi = (-4)^\mathcal{N}\,f\, \bar \Phi, 
\end{eqnarray}
implying again the fact that the highest component of the superfield contains the supersymmetry-breaking scale.
The Lagrangian for $\Phi$ can be written as 
\begin{eqnarray}
	\label{NRocek}
	{\cal L} =  -f \int d^{2\mathcal{N}} \theta \, \Phi .
\end{eqnarray} 
The important constraint in \eqref{phi2N} is only the last one.
The reader can find the demonstration in appendix \ref{appendixA}, where the constraint is explicitly solved in superspace.

Consider therefore a chiral superfield 
\begin{eqnarray}
	\label{chiralN}
	\bar D_{I \dot \alpha} {\cal X}= 0.
\end{eqnarray}
Imposing the constraint 
\begin{equation}
	\label{constrN}
	{\cal X} (D^{I\neq J})^{2\mathcal{N}-2}D^J_\alpha  {\cal X} =0
\end{equation}
and solving it in superspace, we obtain the unique solution
\begin{equation}
	\label{solXN}
	\mathcal{X} = \left(\frac{1}{2}\right)^\mathcal{N} \frac{(G_1^{\alpha} G_{1\alpha})(G_2^{\alpha} G_{2\alpha})\ldots(G_{\mathcal{N}}^\alpha G_{\mathcal{N} \alpha})}{\mathcal{F}^{2\mathcal{N}-1}}\equiv \left(\frac{1}{2}\right)^\mathcal{N} \frac{(G_I)^{2\mathcal{N}}}{\mathcal{F}^{2\mathcal{N}-1}}.
\end{equation}
In this formula the generic ${\cal N}$ superfield ${\cal F}$ is defined such that
\begin{eqnarray}
	\label{auxF_N}
	F={\cal F}|_{\theta^I=0} = \frac{1}{(-4)^\mathcal{N}} (D^I)^{2\mathcal{N}} {\cal X} |_{\theta^I=0} \, ,  
\end{eqnarray}
is the complex scalar auxiliary field, while the superfields $G_{I \alpha}$ are defined such that
\begin{eqnarray}
	\label{GoldN}
	g_{I\alpha}=G_{I\alpha} |_{\theta^I=0} = \frac{1}{\sqrt 2 (-4)^{\mathcal{N} -1}} (D^{J \ne I})^{2\mathcal{N}-2} D^I_\alpha {\cal X} |_{\theta^I=0}  \, , 
\end{eqnarray}
are the goldstini.
Notice that, as a consequence of \eqref{solXN}, ${\cal X}$ satisfies the nilpotency constraint 
\begin{equation}
	{\cal X}^2 = 0 
\end{equation}
and as proved in appendix \ref{appendixA}, it contains only the goldstini and the auxiliary field.
The Lagrangian giving the supersymmetry breaking sector is now
\begin{eqnarray}
	\label{lagrangianNX} 
	{\cal L} = \int d^{4 \mathcal{N}} \theta  \, {\cal X} \bar {\cal X} 
	+ \left( f \int d^{2\mathcal{N}} \theta \, {\cal X}  + c.c. \right) 
\end{eqnarray}
and ${ F}$ has to be integrated out to find a theory including only the goldstini.

In appendix \ref{appendixB} the Lagrangian \eqref{lagrangianNX} is proved to be equivalent to the other Lagrangians for ${\cal N}$ goldstini constructed with different methods.
This demonstrates that our results cover effectively all known formalisms.

\subsection{${\cal N}=1$ superspace}

We are now in the process of formulating the previous results in ${\cal N}=1$ language, which is the most useful one for practical applications.
As a first step we explicitly break the $\textup{U}({\cal N})_R$ covariance by splitting the set of superspace derivatives as
\begin{eqnarray}
	\label{Dsplit}
	D^I_\alpha \rightarrow \{D_\alpha \, , \, D^i_\alpha \},
\end{eqnarray}
where $i$ takes values from $1$ to $\mathcal{N}-1$.

The superfield $X$ of the $\mathcal{N}=1$ $X,Y$ system is\footnote{Since we omit the $\theta$-projections in the definition of $X$, strictly speaking this object lives in the full ${\cal N}$ superspace.
The same observation applies also to $W_{i \alpha}$ in \eqref{Wi}. 
We nevertheless keep refer to them as ${\cal N}=1$ superfields, because this is the role they have been introduced for. 
The $X$, $Y_i$ appearing in \eqref{NNLL} and thereafter are, in any case, properly projected ${\cal N}=1$ superfields. 
We are confident that at this stage of the discussion the reader is going to avoid any type of confusion, even with this little abuse of notation.}
\begin{eqnarray}
	X = \frac{1}{(-4)^{\mathcal{N}-1}}(D^i)^{2\mathcal{N}-2} {\cal X} 
	\, , 
	\qquad \text{such that} \qquad X^2=0,
\end{eqnarray}
and it contains the auxiliary field $F$ and the goldstino of the first supersymmetry $i.e.$, in our conventions, the one related to the first superspace derivative in \eqref{Dsplit} 
\begin{equation}
	g_\alpha=G_\alpha|_{\theta^I=0} =\frac{1}{\sqrt{2}}D_\alpha X|_{\theta^I=0}.
\end{equation}
The other goldstini occupy the highest components of the $\mathcal{N}=1$ superfields 
\begin{eqnarray}
	\label{Wi}
	W_{i\alpha} = \frac{1}{\sqrt 2 (-4)^{\mathcal{N}-2}} (D^{j \ne i})^{2\mathcal{N}-4} D^i_\alpha  {\cal X} , 
\end{eqnarray} 
satisfying the constraints 
\begin{eqnarray}
	W_{i\alpha} = X \frac{D^2 W_{i\alpha} }{D^2 X} .
\end{eqnarray}
In particular in the $\mathcal{N}=1$ language they are described by the superfields 
\begin{equation}
	\qquad G_{i\alpha} = -\frac{1}{4}D^2 W_{i\alpha}.
\end{equation}
The expression of the solution \eqref{solXN} in terms of the goldstini and of the auxiliary field is 
\begin{equation}
	\label{solXN,1}
	\mathcal{X}=2^{5-5{\cal N}}X\frac{\prod_{i=1}^{\mathcal{N}-1}\left(D^2 W_{i\alpha}\right)^2}{\mathcal{F}^{2\mathcal{N}-2}}.
\end{equation}

The lower components of ${\cal X}$ are constrained $\mathcal{N}=1$ superfields and, similarly to the $S$ of the ${\cal N}=2$ case, they are removed from the spectrum in terms of $X$ and  $W_{i \alpha}$.
In particular, they organize in representations of the group $\textup{U}(\mathcal{N}-1)$, acting as a flavour symmetry after the breaking
\begin{equation}
	\textup{U}(\mathcal{N})_R \longrightarrow U(1)_R \times \textup{U}(\mathcal{N}-1).
\end{equation}
To deal with all the possible components of ${\cal X}$ in the general ${\cal N}$ case we proceed as follows.
The generic component can have a number $p$ of fermionic indices, $0\leq p \leq {2{\cal N}}$, and some of them can be contracted in pairs.
An efficient way to handle them consists in distinguishing  between the indices in the set $M_1=\{i_1,\ldots,i_k\}$ which are all different, and the ones in $M_2=\{j_1,\ldots,j_l\}$ which are equal two by two.
In this picture therefore we spilt $p=k+l$ and we construct the general expression 
\begin{equation}
	\label{Sdef}
	\left(S^{\,\,i_1i_2\ldots i_k,\, j_1j_2\ldots j_l}_{(k+l)}\right)_{\alpha_1\alpha_2\ldots\alpha_k,\,\beta_1\beta_2\ldots \beta_l}=\left(\frac{1}{\sqrt{2}}\right)^{k+l} D_{\alpha_1}^{i_1} D_{\alpha_2}^{i_2}\ldots D_{\alpha_k}^{i_k}\,D_{\beta_1}^{j_1} D_{\beta_2}^{j_2}\ldots D_{\beta_l}^{j_l}\,\mathcal{X}|_{\theta^I=0}.
\end{equation}

We conclude this section writing down the Lagrangian of the general ${\cal N}$ theory in the three ${\cal N}=1$ constrained superfield systems we have been discussing in the work, namely the $X,W$ system, the $X,Y$ system and the $X,H$ system.
The $\mathcal{N}=1$ Lagrangian in terms of the superfields $X$ and $W_{i\alpha}$ is of the form 
\begin{eqnarray}
	\label{NNLL}
	{\cal L} = 
	\int d^4 \theta \left( X \bar X 
	+ i \sum_{i} \partial_m W_i^{\alpha} \sigma^m_{\alpha \dot \alpha} \bar W^{i \dot \alpha}  \right) 
	+ f \left( \int d^2 \theta X + c.c. \right) 
	+ \ldots  ,
\end{eqnarray}
where dots stand for higher order terms, containing the $S$ components, which are essential for the non-linear realization of the $\mathcal{N}$ supersymmetries.
Mimicking the discussion of the $\mathcal{N}=2$ theory, we can define the $Y_i$ superfields such that
\begin{eqnarray}
	\label{newYi}
	\begin{split}
		XY_i &= 0 , 
		\\
		\bar X D^2 Y_i  &= 0,  
	\end{split}
\end{eqnarray}
where $X^2=0$ is always understood.
These superfields $Y_i$ contain only fermions, which are going to be the goldstini of the broken supersymmetries.
Using the expression
\begin{eqnarray}
	\label{WiofYi}
	\begin{split}
		W_{i\beta} = & 2\sqrt 2 \frac{X D_\beta Y_i}{D^2 X} 
		+ 16 \sqrt 2 i X \frac{D^\rho Y_i}{D^2 X} \frac{\bar D^{\dot \rho} \bar X}{\bar D^2 \bar X} \partial_{\rho \dot \rho} \left( \frac{D_\beta X}{D^2 X} \right) -
		\\[0.15cm]
		& - 128 \sqrt 2 X \frac{D^\sigma Y_i}{D^2 X} \frac{\bar D^{\dot \sigma} \bar X}{\bar D^2 \bar X} \partial_{\sigma \dot \sigma} \left( \frac{D^\rho X}{D^2 X} \right) 
		\frac{\bar D^{\dot \rho} \bar X}{\bar D^2 \bar X} \partial_{\rho \dot \rho} \left( \frac{D_\beta X}{D^2 X} \right) , 
	\end{split}
\end{eqnarray} 
we can rewrite the Lagrangian \eqref{NNLL} in terms of the superfields $X$ and $Y_i$.
The lower superfields $S_{(p)}$ now take the form 
\begin{equation}
	\label{SNN}
	\begin{aligned}
		(S^{\,\,i_1i_2\ldots i_k,\, j_1j_2\ldots j_l}_{(p)} & )_{\alpha_1\alpha_2\ldots\alpha_k,\,\beta_1\beta_2\ldots \beta_l}=2^{2-2{\cal N}+\frac{3k}{2}-l}\epsilon_{\beta_1\beta_2}\ldots\epsilon_{\beta_{l-1}\beta_l} X\times                                                               \\[0.15cm]
		                                                    & \times \frac{(D_{\rho_1}Y_{i_1}Z^{\rho_1}_{\alpha_1})(D_{\rho_2}Y_{i_2}Z^{\rho_2}_{\alpha_2})\ldots (D_{\rho_k}Y_{i_k}Z^{\rho_k}_{\alpha_k})}{\mathcal{F}^{2{\cal N}-2-k-l}}(D_{\sigma}Y_{i\notin M}Z^{\sigma})^{2{\cal N}-2-2k-l}, 
	\end{aligned}
\end{equation}
where $p=k+l$, $M=M_1+M_2$ and $Z^\rho_\alpha$ is an expression containing only $X$ 
\begin{equation}
	\begin{aligned}
		Z^\rho_\alpha = \delta^\rho_\alpha + 8i \epsilon^{\beta \rho}\frac{\bar D^{\dot\beta} \bar X}{\bar D^2 \bar X}\partial_{\beta \dot\beta}\left(\frac{D_\alpha X}{D^2 X}\right)-64 \epsilon^{\sigma \rho}\frac{\bar D^{\dot\sigma}\bar X}{\bar D^2 \bar X}\partial_{\sigma \dot\sigma}\left(\frac{D^\beta X}{D^2 X}\right) \frac{\bar D^{\dot \beta}\bar X}{\bar D^2 \bar X}\partial_{\beta \dot \beta}\left(\frac{D_\alpha X}{D^2 X}\right) .
	\end{aligned}
\end{equation}
This proves in fact that the only independent superfields are $X$ and the $Y_i$, while all the other superfields are given in terms of them as a consequence of equation \eqref{SNN}.
The expression of the Lagrangian in the $X,Y_i$ system is 
\begin{eqnarray}
	{\cal L}  = \int d^4 \theta \left( X \bar X + \sum_{i} Y_i \bar Y^i [1 + {\cal A} ] \right) 
	+ f \left( \int d^2 \theta X + c.c. \right) 
	+ {\cal L}_\text{HO}  , 
\end{eqnarray}
where ${\cal A}$ is defined as in \eqref{Aterm} and ${\cal L}_\text{HO} $ stands for higher order terms, making the theory invariant under the non-linearly realized additional supersymmetries.

Let us give a rather compact expression for the Lagrangian, which can be useful in practical calculations.
With a bit of reordering the complete theory can be recast into the form
\begin{equation}
	\label{Lcompact}
	{\cal L}=\int d^4\theta \,\sum_{l=0}^{{\cal N}-1}\,\sum_{k=0}^{{\cal N}-1}L^{({\cal N})}_{k,l}+ f \left( \int d^2 \theta X + c.c. \right) , 
\end{equation}
with
\begin{equation}
	\begin{aligned}
		L^{({\cal N})}_{k,l} & =\sum_{i_1,\ldots,i_k=1}^{{\cal N}-1}\,\,\,\sum_{j_l>j_{l-2}>\ldots>j_2=1}^{{\cal N}-1}(-i)^k\,2^{-l}\,  (S^{i_1\ldots i_k,j_1\ldots j_l})^{\alpha_1\ldots \alpha_k,\,\beta_2\beta_4\ldots\beta_l}_{\phantom{\alpha_1\ldots \alpha_k,\beta_2\beta}\beta_2\beta_4\ldots\beta_l}\times                                                                      \\
		                    & \times\partial_{\alpha_1\dot{\alpha}_1}\ldots\partial_{\alpha_k\dot{\alpha}_k}(\partial^2)^{{\cal N}-1-k-\frac{l}{2}}\,(\bar{S}_{i_1\ldots i_k,j_1\ldots j_l})^{\dot{\alpha}_1\ldots \dot{\alpha}_k,\,\phantom{\beta_2}\dot{\beta}_2\dot{\beta}_4\ldots\dot{\beta}_l}_{\phantom{\alpha_1\ldots \alpha_k}\,\dot{\beta}_2\dot{\beta}_4\ldots\dot{\beta}_l} 
	\end{aligned}
\end{equation} 
and where $\bar{S}$ is defined with the fermionic indices in the opposite order 
\begin{equation}
	(\bar{S}_{\,\,i_1i_2\ldots i_k,\, j_1j_2\ldots j_l}^{(k+l)})_{\dot{\alpha}_1\dot{\alpha}_2\ldots\dot{\alpha}_k,\,\dot{\beta}_1\dot{\beta}_2\ldots \dot{\beta}_l}=\left(\frac{1}{\sqrt{2}}\right)^{k+l} \bar{D}_{\dot{\alpha}_k}^{i_k} \bar{D}_{\dot{\alpha}_{k-1}}^{i_{k-1}}\ldots \bar{D}_{\dot{\alpha}_1}^{i_1}\,\bar{D}_{\dot{\beta}_l}^{j_l} \bar{D}_{\dot{\beta}_{l-1}}^{j_{l-1}}\ldots \bar{D}_{\dot{\beta}_1}^{j_1}\,\bar{\mathcal{X}}| .
\end{equation}
With this formula one can effectively extract the contribution of the desired derivative order without necessarily compute the whole Lagrangian.
As a check, in the ${\cal N}=2$ case where $L^{(2)}_{1,2}$ vanishes, we have 
\begin{equation}
	\begin{aligned}
		{\cal L} & = \int d^4\theta \left(L^{(2)}_{0,0}+L^{(2)}_{1,0}+L^{(2)}_{0,2}\right)+f\left(\int d^2\theta X + c.c.\right) =                                                \\
		        & =\int d^4\theta \left(S \partial^2 \bar{S}-iW^\alpha\partial_{\alpha\dot{\alpha}}\bar{W}^{\dot{\alpha}}+ X\bar{X}\right)+f\left(\int d^2\theta X + c.c.\right) 
	\end{aligned}
\end{equation}
and it matches exactly with \eqref{N2LL}.

Finally, an alternative way to express the theory is in terms of the chiral superfields $X$ and the $H^i_{\dot \alpha}$ satisfying
\begin{equation}
	\begin{aligned}
		X^2                                       & =0 ,  
		\\ 
		\bar D_{\dot \alpha} (X \bar H_{i\beta} ) & = 0 .
	\end{aligned}
\end{equation}
The Lagrangian is always given by formula \eqref{Lcompact}, but now 
\begin{equation}
	\begin{aligned}
		S^{\,\,i_1i_2\ldots i_k,\, j_1j_2\ldots j_l}_{(k+l)\alpha_1\alpha_2\ldots\alpha_k,\,\beta_1\beta_2\ldots \beta_l} & =      
		(-4)^{2\mathcal{N}-2-k-l} 2^{5-5{\cal N}+3k+\frac{5}{2}l} \epsilon_{\beta_1\beta_2}\ldots\epsilon_{\beta_{l-1}\beta_l}X\times
		\\[0.1cm]
		                                                                                                                  & \times 
		\bar H_{i_1 \alpha_1}  
		\bar H_{i_2 \alpha_2} 
		\ldots 
		\bar H_{i_k \alpha_k} 
		(\bar H_{i\notin M})^{2{\cal N}-2-2k-l}.
	\end{aligned}
\end{equation}

\begin{table}[h]
	\label{tabb3} 
	\begin{center}
		$$
		\begin{array}{ccccc}
			\hline 
			\text{\bf SUSY }              & {\cal N}=1 & {\cal N}=2    & {\cal N}=4                 & \text{generic ${\cal N}$}                \\
			\hline 
			\text{Goldstini Superfields}  & X          & X ,  W        & X , W_1 ,  W_2 , W_3       & X ,  W_1 ,  W_2 , \ldots, W_{{\cal N}-1} \\
			\hline
			\text{Goldstini Superfields}  & X          & X ,  H        & X , H^1 ,  H^2 , H^3       & X ,  H^1 ,  H^2 , \ldots, H^{{\cal N}-1} \\
			\hline
			\text{Goldstini Superfields}  & X          & X ,  Y        & X , Y_1 ,  Y_2 , Y_3       & X ,  Y_1 ,  Y_2 , \ldots, Y_{{\cal N}-1} \\
			\hline
			\text{Eliminated Superfields} & -          & S^{(0)}       & S^{(0)},  S^{(1)}, S^{(2)} &                                         
			S^{(0)},S^{(1)},\,\ldots\,,S^{(2{\cal N}-4)}\\
			\hline
			\text{Residual Flavor Group}  & -          & \textup{U}(1) & \textup{U}(3)              & \textup{U}({\cal N}-1)                  \\
			\hline
		\end{array}
		$$
	\end{center}
	\caption{\small The ${\cal N}=1$ chiral superfields content of a minimal ${\cal N}$ goldstini theory.
The shorthand notation $S^{(p)}$ indicates all the possible components \eqref{Sdef} with $p$ fermionic indices contracted in all the possible ways.} 
\end{table}

\section{Discussion} 
\label{sec:discussion}

In this work we studied spontaneous supersymmetry breaking in four dimensions, involving a generic number ${\cal N}$ of supersymmetries.
We focused on the supersymmetry breaking sector of the theory and we identified its structure in all known formulations, proving explicitly their equivalence.
Our results are presented in the superspace setup, which is the preferred framework to study supersymmetric theories.
In particular, we identified the properties of the goldstino superfields among the various formulations and the constraints they satisfy.
Thanks to the aforementioned equivalence, one may use our results either in the language of ${\cal N}$ supersymmetries, or in the standard language of ${\cal N}=1$ supersymmetry, depending on the application one is interested in and knowing directly the relation to the other formulations.

Our results lead the way to describe low energy theories with ${\cal N}$ spontaneously broken supersymmetries, 
in the setup of constrained superfields.
In this language, one can have various components of the matter sector that are going to be removed from the spectrum by imposing appropriate constraints of the form  described in \cite{Dall'Agata:2016yof}.
Within the same setup one can also study properties of  theories where the goldstini originate from some underlying strong dynamics.
The ${\cal N}$ non-linear realizations will be then violated by possible couplings to the Higgs field via the Yukawa couplings.
Considering $e.g.$ the Standard Model, the pseudo-goldstini can be interpreted as its matter content.
In this way one has a dynamical scheme exploiting supersymmetry to produce (almost) massless fermions.
Ignoring the gauge and Higgs sectors, from our results one can read the restrictions imposed on the various interactions between the quarks and leptons.
For example it can be seen directly that terms of the form $\int d^2 \theta \, Y^2$ are not allowed.
The obvious next step is the analysis of the couplings of this sector to vector fields.
Following the approach presented here, one could try either coupling the ${\cal N}$-goldstini sector to standard gauge multiplets (up to ${\cal N}=4$) or to constrained multiplets whose surviving degrees of freedom are gauge fields.
Both avenues require substantial work that goes beyond the scope of the present paper.

Non-linear realizations of supersymmetric theories have been recently revisited in several different contexts and we believe our work will offer new directions for further developments.
It would be indeed interesting to perform a systematic study for general $\mathcal{N}$ broken supersymmetries
within supergravity, building on the work of \cite{Dudas:2015eha}-\cite{Farakos:2016hly}.
This might also help finding new interesting scenarios for inflationary cosmology, providing new
models with non-linear realizations of supersymmetry along the lines of \cite{Antoniadis:2014oya}-\cite{Ferrara:2016ajl}.
Finally, while one might suspect that ultraviolet completions within string theory may exist only for special values of ${\mathcal N}$, it would be interesting to understand which string compactifications may give rise to such non-linear realizations \cite{Bergshoeff:2015jxa}-\cite{Kallosh:2016aep}.

\section*{Acknowledgements}

We thank F. Riva, F. Zwirner and especially R. Rattazzi for discussions.
This work was  supported in part by the MIUR grant RBFR10QS5J (STaFI) and by the Padova University Project CPDA119349.

\appendix

\section{Solution and consistency of the constraint for ${\cal N}$ goldstini}
\label{appendixA}

In this appendix we explicitly derive the solution \eqref{solXN} from the constraint \eqref{constrN}.
We show then that the solution is chiral and that it contains only the goldstini and the auxiliary field which breaks supersymmetry.

To derive the solution \eqref{solXN}  notice first that the goldstini $G_{I\alpha}$ satisfy the property
\begin{equation}
	\label{goldpropN}
	D^I_\beta G_{J\alpha} = \sqrt{2}\mathcal{F}\epsilon_{\alpha\beta}\delta^{I}_{J}.
\end{equation}
Using \eqref{GoldN}, the constraint \eqref{constrN} can be expressed in the more compact form
\begin{equation}
	\label{constrN2}
	{\cal X} G_{J\alpha} = 0.
\end{equation}
Consider now a particular fixed index $\hat{J}$\footnote{In the following every hatted index is not summed.}.
Dividing \eqref{constrN2} by $\mathcal{F}$, acting with $D^{\hat{J}\alpha }$ and using the property \eqref{goldpropN} we have
\begin{equation}
	D^{\hat{J}\alpha}\left(\frac{\mathcal{X}G_{\hat{J} \alpha}}{\mathcal{F}}\right)=D^{\hat{J}\alpha}\mathcal{X}\frac{G_{\hat{J} \alpha}}{\mathcal{F}}+\mathcal{X}\frac{D^{\hat{J}\alpha} G_{\hat{J} \alpha}}{\mathcal{F}} = 0
\end{equation}  
and thus
\begin{equation}
	\label{xproof1}
	\mathcal{X}=\frac{D^{\hat{J}\alpha}\mathcal{X} G_{\hat{J} \alpha}}{2\sqrt{2}\mathcal{F}}.
\end{equation}
Acting again with the same $D^{\hat{J}\beta}$ we get
\begin{equation}
	D^{\hat{J}\beta}\mathcal{X}=\frac{D^{\hat{J}\beta}D^{\hat{J}\alpha}\mathcal{X}G_{\hat{J} \alpha}-D^{\hat{J}\alpha}\mathcal{X}D^{\hat{J}\beta}G_{\hat{J}\alpha}}{2\sqrt{2}\mathcal{F}}=-\frac{1}{4}(D^{\hat{J}})^2\mathcal{X}\frac{G_{\hat{J}}^\beta}{\sqrt{2}\mathcal{F}}+\frac{1}{2}D^{\hat{J}\beta}\mathcal{X}
\end{equation}
and therefore
\begin{equation}
	\label{xproof2}
	D^{\hat{J}\alpha}\mathcal{X}=-\frac{(D^{\hat{J}})^2\mathcal{X}\,G_{\hat{J}}^{\alpha}}{2\sqrt{2}\mathcal{F}} .
\end{equation}
Inserting \eqref{xproof2} in \eqref{xproof1}, we get
\begin{equation}
	\mathcal{X}=\frac{D^{\hat{J}\alpha}\mathcal{X}G_{\hat{J} \alpha}}{2\sqrt{2}\mathcal{F}}=-\frac{(D^{\hat{J}})^2\mathcal{X}G_{\hat{J}}^{\alpha}G_{\hat{J}\alpha}}{8\mathcal{F}^2},
\end{equation}
$i.e.$
\begin{equation}
	\mathcal{X}=\frac{-\frac{1}{4}(D^{\hat{J}})^2\mathcal{X}}{\mathcal{F}}\frac{(G_{\hat{J}})^2}{2\mathcal{F}}.
\end{equation}
This expression in particular shows that $\mathcal{X}$ is proportional to any squared golstino.
Consider now two fixed indices $\hat{J}\neq \hat{K}$ (if $\hat{J}=\hat{K}$ the expression vanishes trivially) 
\begin{equation}
	\begin{aligned}
		\mathcal{X} & = \frac{-\frac{1}{4}(D^{\hat{J}})^2\mathcal{X}}{\mathcal{F}} \frac{(G_{\hat{J}})^2}{2\mathcal{F}}, \\
		\mathcal{X} & = \frac{-\frac{1}{4}(D^{\hat{K}})^2\mathcal{X}}{\mathcal{F}} \frac{(G_{\hat{K}})^2}{2\mathcal{F}}.
	\end{aligned}
\end{equation}
Inserting the second expression into the first we have
\begin{equation}
	\mathcal{X} = \frac{\left[-\frac{1}{4}(D^{\hat{J}})^2\right]}{\mathcal{F}}\frac{\left[-\frac{1}{4}(D^{\hat{K}})^2\mathcal{X}\right]}{\mathcal{F}} \frac{(G_{\hat{K}})^2}{2\mathcal{F}} \frac{(G_{\hat{J}})^2}{2\mathcal{F}}
\end{equation}
and, by repeating $\mathcal{N}$ times the trick, we obtain using \eqref{auxF_N} 
\begin{equation}
	\mathcal{X} = \frac{\left(-\frac{1}{4}\right)^\mathcal{N}(D^I)^{2\mathcal{N}}\mathcal{X}}{\mathcal{F}^{\mathcal{N}}}\frac{(G_I)^{2\mathcal{N}}}{(2\mathcal{F})^\mathcal{N}}=\left(\frac{1}{2}\right)^{\mathcal{N}}\frac{(G_I)^{2\mathcal{N}}}{{\mathcal{F}}^{2\mathcal{N}-1}},
\end{equation}
which is our result \eqref{solXN}.

As a first consistency check we prove that our solution \eqref{solXN} is chiral, namely that
\begin{equation}
	\bar D_{\dot\alpha}^K \left[\left(\frac{1}{2}\right)^\mathcal{N}\frac{(G_I)^{2\mathcal{N}}}{\mathcal{F}^{2\mathcal{N}-1}}\right]=0.
\end{equation}
We first notice that
\begin{equation}
	\bar D_{\dot\alpha}^K\mathcal{F}=-\sqrt{2}i\partial_{\beta\dot\alpha}G_K^{\beta}.
\end{equation}
Since the action of the covariant derivatives $\bar{D}$ on the goldstini is
\begin{equation}
	\bar D_{\dot\alpha}^K G_{I \alpha}=\frac{1}{\sqrt{2}(-4)^{\mathcal{N}-1}}\left[(D^{J\neq \hat{I}\neq \hat{K}})^{2\mathcal{N}-4}\left(4i\partial_{\beta\dot{\alpha}}D^{\hat{K}\,\beta}D^{I\neq \hat{K}}_\alpha \mathcal{X}\right)-2i\partial_{\alpha\dot\alpha}(D^{J\neq \hat{I}})^{2\mathcal{N}-2}\mathcal{X}\delta^{\hat{I}}_{\hat{K}}\right],
\end{equation}
where we distinguished the contribution from the $I=K$ and the $I\neq K$ case, we need also the expressions
\begin{equation}
	(D^{J\neq\hat{I}})^{2\mathcal{N}-2}\mathcal{X}=-2(-4)^{\mathcal{N}-2}\frac{(G_{\hat{I}})^2}{\mathcal{F}}
\end{equation}
and
\begin{equation}
	D^{\hat{I}}_\alpha D^{\hat{K}}_\beta(D^{J\neq \hat{I}\neq\hat{K}})^{2\mathcal{N}-4}\mathcal{X}=-\frac{(-4)^{\mathcal{N}-1}G_{\hat{I} \alpha} G_{\hat{K}\beta}}{2\mathcal{F}} .
\end{equation}
We have now all the ingredients to calculate
\begin{equation}
	\begin{aligned}
		\bar D_{\dot\alpha}^K \left[\left(\frac{1}{2}\right)^\mathcal{N}\frac{(G_I)^{2\mathcal{N}}}{\mathcal{F}^{2\mathcal{N}-1}}\right] &    		=
		\frac{1}{2^\mathcal{N}}
		\left[
		-\frac{2}{\mathcal{F}^{2\mathcal{N}-1}}
		\sum_{\hat{I}=1}^\mathcal{N}
		(G_{J\neq \hat{I}})^{2\mathcal{N}-2}G_{\hat{I}}^{\alpha}\bar D_{\dot\alpha}^KG_{\hat{I} \alpha}-\frac{(2\mathcal{N}-1)(G_{I})^{2\mathcal{N}}}{\mathcal{F}^{2\mathcal{N}}}\bar D_{\dot\alpha}^K\mathcal{F}
		\right]
		=
		\\
		                                                                                                                                 & =\frac{1}{2^\mathcal{N}}                                                             		\bigg[
		\frac{2\sqrt{2}i}{\mathcal{F}^{2\mathcal{N}-1}}\sum_{\hat{I}=1}^{\mathcal{N}}(G_{J\neq \hat{I}})^{2\mathcal{N}-2}G_{\hat{I}}^{\alpha}\partial_{\beta\dot\alpha}\left(\frac{G_{\hat{K}}^{\beta}G_{\hat{I}\neq\hat{K} \, \alpha}}{\mathcal{F}}\right)+ 
		\\
		                                                                                                                                 & +\frac{\sqrt{2}i}{\mathcal{F}^{2\mathcal{N}-1}}(G_{J\neq\hat{K}})^{2\mathcal{N}-2}G_{\hat{K}}^{\alpha}\partial_{\alpha\dot\alpha}\left(\frac{(G_{\hat{K}})^2}{\mathcal{F}}\right)+\sqrt{2}i(2\mathcal{N}-1)\frac{(G_I)^{2\mathcal{N}}}{\mathcal{F}^{2\mathcal{N}}}\partial_{\beta\dot\alpha}G_{K}^{\beta}\bigg]                                  
		=
		\\
		                                                                                                                                 & =\frac{\sqrt{2}i(\mathcal{N}-1)}{2^{\mathcal{N}-1}\mathcal{F}^{2\mathcal{N}}}(G_I)^{2\mathcal{N}}\partial_{\beta\dot\alpha}G_K^{\beta}-\frac{\sqrt{2}i}{2^{\mathcal{N}-1}\mathcal{F}^{2\mathcal{N}-1}}\sum_{\hat{I}=1,\, \hat{I}\neq\hat{K}}^\mathcal{N}\frac{(G_{J})^{2\mathcal{N}}\partial_{\beta\dot\alpha}G_{\hat{K}}^{\beta}}{\mathcal{F}}= 
		\\
		                                                                                                                                 & =\frac{\sqrt{2}i(\mathcal{N}-1)}{2^{\mathcal{N}-1}\mathcal{F}^{2\mathcal{N}}}(G_I)^{2\mathcal{N}}\partial_{\beta\dot\alpha}G_K^{\beta}-\frac{\sqrt{2}i}{2^{\mathcal{N}-1}\mathcal{F}^{2\mathcal{N}-1}}\frac{(G_{I})^{2\mathcal{N}}\partial_{\beta\dot\alpha}G_K^{\beta}}{\mathcal{F}}(\mathcal{N}-1)=                                            
		\\
		                                                                                                                                 & =0.                                                                                  	\end{aligned}
\end{equation}
This proves that the solution \eqref{solXN} is chiral.

We show finally that the solution \eqref{solXN} contains only goldstini and the auxiliary field breaking supersymmetry.
We verify first \eqref{auxF_N}, namely
\begin{equation}
	\mathcal{F}=\frac{1}{(-4)^\mathcal{N}}(D^I)^{2\mathcal{N}}\mathcal{X}.
\end{equation}
The action of $(D^I)^{2\mathcal{N}}$ on $(G_{I})^{2\mathcal{N}}$ is 
\begin{equation}
	\begin{aligned}
		(D^I)^{2\mathcal{N}}(G_{I})^{2\mathcal{N}} & =(D^{I\neq\hat{J}})^{2\mathcal{N}-2}(G_{I\neq\hat{J}})^{2\mathcal{N}-2}D^{\hat{J}\beta}(-2G_{\hat{J}}^{\alpha}D^{\hat{J}}_\beta G_{\hat{J} \alpha})= 
		\\
		                                           & =(D^{I\neq\hat{J}})^{2\mathcal{N}-2}(G_{I\neq\hat{J}})^{2\mathcal{N}-2}(-2D^{\hat{J}\beta}G_{\hat{J}}^{\alpha}D^{\hat{J}}_\beta G_{\hat{J} \alpha})= 
		\\
		                                           & =(D^{I\neq\hat{J}})^{2\mathcal{N}-2}(G_{I\neq\hat{J}})^{2\mathcal{N}-2}(-2(2\mathcal{F})^2)=                                                         
		\\
		                                           & =\ldots=                                                                                                                                             
		\\
		                                           & =(-2)^\mathcal{N}(2\mathcal{F})^{2\mathcal{N}},                                                                                                      
	\end{aligned}
\end{equation}
where dots mean we have repeated $\mathcal{N}$ times the same steps.
We can now calculate
\begin{equation}
	\begin{aligned}
		\frac{1}{(-4)^\mathcal{N}}(D^I)^{2\mathcal{N}}\mathcal{X} & = \frac{1}{(-4)^\mathcal{N}}(D^{I})^{2\mathcal{N}}\left[\left(\frac{1}{2}\right)^\mathcal{N}\frac{(G_I)^{2\mathcal{N}}}{\mathcal{F}^{2\mathcal{N}-1}}\right]=             
		\\
		                                                          & =\frac{1}{(-4)^\mathcal{N}}\left(\frac{1}{2}\right)^\mathcal{N}\frac{1}{\mathcal{F}^{2\mathcal{N}-1}}(-2)^\mathcal{N}4^\mathcal{N}\mathcal{F}^{2\mathcal{N}}=\mathcal{F}.
	\end{aligned}
\end{equation}
We verify then \eqref{GoldN}, namely
\begin{equation}
	G_{I \alpha}=\frac{1}{\sqrt{2}(-4)^{\mathcal{N}-1}}(D^{J\neq I})^{2\mathcal{N}-2}D^I_\alpha \mathcal{X}.
\end{equation}
Using the previous result, the action of $(D^{J\neq I})^{2\mathcal{N}-2}D^I_\alpha$ on $(G_I)^{2\mathcal{N}}$ is
\begin{equation}
	\begin{aligned}
		(D^{J\neq I})^{2\mathcal{N}-2}D^I_\alpha (G_K)^{2\mathcal{N}} & =(D^{J\neq I})^{2\mathcal{N}-2}(G_{J\neq I})^{2\mathcal{N}-2}(-2 G_I^{\beta}D^I_\alpha G_{I \beta})= 
		\\
		                                                              & =\sqrt{2}(-2)^{\mathcal{N}-1}(2\mathcal{F})^{2\mathcal{N}-1}G_{I \alpha}.
	\end{aligned}
\end{equation} 
We can now calculate
\begin{equation}
	\begin{aligned}
		\frac{1}{\sqrt{2}(-4)^{\mathcal{N}-1}}(D^{J\neq I})^{2\mathcal{N}-2}D^I_\alpha \mathcal{X} & =\frac{1}{\sqrt{2}(-4)^{\mathcal{N}-1}}(D^{J\neq I})^{2\mathcal{N}-2}D^I_\alpha\left[\left(\frac{1}{2}\right)^\mathcal{N}\frac{(G_I)^{2\mathcal{N}}}{\mathcal{F}^{2\mathcal{N}-1}}\right]= 
		\\
		                                                                                           & =\frac{1}{\sqrt{2}(-4)^{\mathcal{N}-1}}\left(\frac{1}{2}\right)^\mathcal{N}\frac{1}{\mathcal{F}^{2\mathcal{N}-1}}\sqrt{2}(-2)^{\mathcal{N}-1}(2\mathcal{F})^{2\mathcal{N}-1}G_{I \alpha}=  
		\\
		                                                                                           & =G_{I \alpha}.
	\end{aligned}
\end{equation}

\section{Equivalence of the goldstini models in ${\cal N}$ superspace}
\label{appendixB}

In this appendix we explicit show the equivalence among all the known models of goldstini, working directly in ${\cal N}$ superspace.
We start by presenting briefly how to generalize the Volkov--Akulov model in ${\cal N}$ 
superspace with the Samuel--Wess formalism.
This integrates our discussion of section \ref{sec:N=1_goldstino_in_superspace}.

The Lagrangian \eqref{LN} can also be obtained in a geometrical way.
Defining the superspace matrix
\begin{equation}
	\mathbb{A}_{m}^a = \delta_{m}^a-\frac{i}{f^2}\sum_{I}\partial_m \Lambda_I\sigma^a\bar\Lambda^I+\frac{i}{f^2}\sum_I\Lambda_I\sigma^a \partial_m\bar\Lambda^I,
\end{equation} 
where $\Lambda_{I\,\alpha}$ satisfy \eqref{NSWrepr}, one can construct the invariant Lagrangian
\begin{equation}
	\label{LNgeom}
	{\cal L} = -f^2 \det \mathbb{A}_m^a|_{\theta^I=0}.
\end{equation}
The equivalence between \eqref{LN} and \eqref{LNgeom} can be proved in the following way.
Due to the particular form of $\mathbb{A}_m^a$ we have 
\begin{equation}
	\begin{aligned}
		\int d^{4{\cal N}}\theta \Lambda^{2{\cal N}}\bar\Lambda^{2{\cal N}} & =   \int d^{4{\cal N}}\theta \Lambda^{2{\cal N}}\bar\Lambda^{2{\cal N}}\det\mathbb{A}_m^a                                                                   
		\\
		                                                                 & = \frac{1}{(-4)^{2{\cal N}}}(D^I)^{2{\cal N}}(\bar D^J)^{2{\cal N}}\left( \Lambda^{2{\cal N}}\bar\Lambda^{2{\cal N}}\det\mathbb{A}_m^a \right)|_{\theta^I=0}.
	\end{aligned}
\end{equation}
Acting then with the covariant derivatives inside the parenthesis the $\Lambda$ superfields are removed.
In fact for generic numbers $p,q$ of spinor superfields
\begin{equation}
	\label{proof}
	\begin{aligned}
		  & D^{\hat{I}\,\alpha}D^{\hat{I}}_\alpha\left(\Lambda_{I_1\,\alpha_1}\ldots\Lambda_{I_p\,\alpha_p}\Lambda^\beta_{\hat{I}}\Lambda_{\hat{I}\,\beta}\bar{\Lambda}^{J_1}_{\dot{\alpha}_1}\ldots\bar{\Lambda}^{J_q}_{\dot{\alpha}_q}\det \mathbb{A}_m^a\right)=                                                                                                                                                            
		\\
		  & =D^{\hat{I}\alpha}\left[\frac{i}{f}\bar\Lambda^{\hat{I}\,\dot{\rho}}\sigma^a_{\alpha\dot{\rho}}\partial_a\left(\underline{\Lambda}_{p,q}{\Lambda^\beta}_{\hat{I}}\Lambda_{\hat{I}\,\beta}\right)\det \mathbb{A}_m^a+2\Lambda_{\hat{I}\,\alpha}\underline{\Lambda}_{p,q}\det \mathbb{A}_m^a+\Lambda^\beta_{\hat{I}}\Lambda_{\hat{I}\,\beta}{D_\alpha}^{\hat{I}}\det \mathbb{A}_m^a\underline{\Lambda}_{p,q}\right], 
	\end{aligned}
\end{equation}
where hatted indices are fixed, $i.e.$ not summed, $\underline{\Lambda}_{p,q}=\Lambda_{I_1\,\alpha_1}\ldots\Lambda_{I_p\,\alpha_p}\,\bar{\Lambda}^{J_1}_{\dot{\alpha}_1}\ldots\bar{\Lambda}^{J_q}_{\dot{\alpha}_q}$
and where we used the property
\begin{equation}
	D^{\hat{I}}_\alpha\det \mathbb{A}_m^a = \frac{i}{f}\partial_a\left(\sigma^a_{\alpha\dot{\rho}}\bar\Lambda^{\hat{I}\,\dot{\rho}}\det \mathbb{A}_m^a\right).
\end{equation}
Integrating by parts and up to total derivatives, \eqref{proof} reduces to 
\begin{equation}
	\begin{aligned}
		D^{\hat{I}\,\alpha}\left(2\Lambda_{\hat{I}\,\alpha}\underline{\Lambda}_{p,q}\det \mathbb{A}_m^a\right) & =-2  \bigg(\frac{i}{f}\bar\Lambda^{\hat{I}\,\dot{\rho}}\sigma^a_{\alpha\dot{\rho}}\partial_a\left({\Lambda^\alpha}_{\hat{I}}\underline{\Lambda}_{p,q}\right)\det \mathbb{A}_m^a+2 f \underline{\Lambda}_{p,q}\det \mathbb{A}_m^a+ \\
		                                                                                                       & +\frac{i}{f}\partial_a\left(\sigma^a_{\alpha\dot{\rho}}\bar\Lambda^{\hat{I}\,\dot\rho}\det \mathbb{A}_m^a\right){\Lambda^\alpha}_{\hat{I}}\underline{\Lambda}_{p,q}\bigg)=                                                        \\
		                                                                                                       & =-2 f (2\underline{\Lambda}_{p,q}\det \mathbb{A}_m^a) .
	\end{aligned}
\end{equation}
Using this result and specializing to the correct number of fermions and derivatives, one can prove that
\begin{equation}
	\int d^4x d^{4{\cal N}}\theta\Lambda^{2{\cal N}}\bar{\Lambda}^{2{\cal N}} = f^{4 {\cal N} } \int d^4x \det \mathbb{A}_m^a|_{\theta^I=0} .
\end{equation}
We know that, in addition to the Samuel--Wess superfield we have been using so far, 
one can define a second representation 
\begin{eqnarray}
	\label{NSWrepr2}
	\begin{split} 
		D_\alpha^I \Gamma_{J\,\beta} & = f\,\epsilon_{\beta\alpha} \, \delta^I_J , 
		\\[0.1cm]
		\bar D_{I\,\dot \alpha} \Gamma_{J\,\beta} & = - \frac{2i}{f} \, \Gamma^{\rho}_I \, \sigma^m_{\rho \dot \alpha} 
		\partial_m \Gamma_{J\,\beta} ,  
	\end{split}
\end{eqnarray} 
the relation between the two representations being established via 
\begin{equation}
	\Gamma_{I\,\alpha} = - 2 f \frac{ (D^{J \ne I})^{2\mathcal{N}-2} D^I_\alpha \bar{D}^{2 {\cal N}} \left( \Lambda^{2{\cal N}}\bar{\Lambda}^{2{\cal N}} \right) }{
		D^{2 {\cal N}} \bar{D}^{2 {\cal N}} \left( \Lambda^{2{\cal N}}\bar{\Lambda}^{2{\cal N}} \right) 
	} .
\end{equation}
The Samuel--Wess Lagrangian for this model is 
\begin{equation}
	{\cal L} =- \frac{1}{f^{4 {\cal N} -2}} \int d^{4{\cal N}}\theta \, \Gamma^{2{\cal N}}\bar{\Gamma}^{2{\cal N}} 
\end{equation}
and it is equivalent to \eqref{LNgeom} because 
\begin{equation}
	\label{LG}
	\Gamma^{2{\cal N}}\bar{\Gamma}^{2{\cal N}}  = \Lambda^{2{\cal N}}\bar{\Lambda}^{2{\cal N}}  .
\end{equation}
For later convenience, notice that we also have 
\begin{equation}
	\label{XGN}
	\Gamma_{I\,\alpha} =  \frac{f}{\sqrt 2} \, \frac{G_{I\,\alpha}}{ {\cal F} } .
\end{equation} 
We have just proved that the extended Volkov--Akulov model is equivalent to the Samuel--Wess formulation, independently from its representation.
We know that this last formulation, say the Lagrangian \eqref{LN}, is equivalent to the extended Rocek's Lagrangian \eqref{NRocek}, as a consequence of \eqref{Finn}.
There remains to demonstrate therefore the equivalence between the Volkov--Akulov model and the Komargodski--Seiberg realization \eqref{lagrangianNX}.
To this purpose we introduce a new set of superspace derivatives 
\begin{equation}
	\begin{split}
		\Pi^I_\alpha &= D_\alpha^I - \frac{i}{f} \sigma^n_{\alpha \dot \alpha} \bar \Lambda^{I \dot \alpha} \partial_n , 
		\\[0.1cm]
		\bar \Pi_{I \dot \alpha} &= \bar D_{I \dot \alpha} + \frac{i}{f} \Lambda_I^{\alpha} \sigma^n_{\alpha \dot \alpha}  \partial_n ,
	\end{split}
\end{equation}
realizing the algebra 
\begin{equation}
	\begin{split}
		\{ \Pi^I_\alpha , \Pi^J_\beta \}  &= 0 , 
		\\[0.1cm]
		\{ \Pi^I_\alpha , \bar \Pi_{J \dot \beta} \} &= 0 .
	\end{split}
\end{equation} 
From these derivatives we can now build projection operators turning a linear realization into a standard non-linear one.
For a generic superfield $U$ in fact we have 
\begin{equation}
	\begin{split}
		D_\alpha^I \left(  \Pi^{2 {\cal N}} \bar{\Pi}^{2 {\cal N}} U \right) 
		& = \frac{i}{f} \sigma^n_{\alpha \dot \alpha} \bar \Lambda^{I \dot \alpha} \partial_n  \left(  \Pi^{2 {\cal N}} \bar{\Pi}^{2 {\cal N}} U \right) , 
		\\[0.1cm]
		\bar D_{I \dot \alpha} \left(  \Pi^{2 {\cal N}} \bar{\Pi}^{2 {\cal N}} U \right) 
		& = - \frac{i}{f} \Lambda_I^{\alpha} \sigma^n_{\alpha \dot \alpha}  \partial_n   \left(  \Pi^{2 {\cal N}} \bar{\Pi}^{2 {\cal N}} U \right)  , 
	\end{split}
\end{equation} 
implying also
\begin{equation}
	\begin{split}
		D^{I}_\alpha \Big{[} \det \mathbb{A}_m^a \left( \Pi^{2 {\cal N}} \bar{\Pi}^{2 {\cal N}} U \right) \Big{]} 
		&= \frac{i}{f}\partial_a \Big{[}  \sigma^a_{\alpha\dot{\rho}}\bar\Lambda^{I\,\dot{\rho}}\det \mathbb{A}_m^a  
		\left(  \Pi^{2 {\cal N}} \bar{\Pi}^{2 {\cal N}} U \right) \Big{]} , 
		\\[0.1cm]
		\bar D_{I \dot \alpha} \Big{[} \det \mathbb{A}_m^a \left( \Pi^{2 {\cal N}} \bar{\Pi}^{2 {\cal N}} U \right) \Big{]} 
		&= - \frac{i}{f}\partial_a \Big{[} \Lambda_I^{\alpha} \sigma^n_{\alpha \dot \alpha}   \det \mathbb{A}_m^a  
		\left(  \Pi^{2 {\cal N}} \bar{\Pi}^{2 {\cal N}} U \right) \Big{]}  , 
	\end{split}
\end{equation} 
for bosonic $U$.

Using \eqref{XGN}, the Lagrangian \eqref{lagrangianNX} can be written as 
\begin{equation}
	\label{lagrangianNXproof}
	{\cal L} = \frac{1}{f^{4 {\cal N}}} \int d^{4 \mathcal{N}} \theta  \, \Gamma^{2{\cal N}}\bar{\Gamma}^{2{\cal N}}   
	\left( {\cal F} \bar{\cal F} + f {\cal F}  + f \bar{\cal F}   \right) .
\end{equation}
As a consequence of \eqref{LG} and of the property 
\begin{equation} 
	\Lambda^{2{\cal N}}\bar{\Lambda}^{2{\cal N}}  {\cal F} 
	= \Lambda^{2{\cal N}}\bar{\Lambda}^{2{\cal N}}  \mathbb{F}, 
\end{equation}
where 
\begin{equation} 
	\mathbb{F} = \frac{1}{(16 f^2)^\mathcal{N}} \Pi^{2 {\cal N}} \bar{\Pi}^{2 {\cal N}} \left( {\cal X} \, \bar \Gamma^{2{\cal N}} \right), 
\end{equation} 
\eqref{lagrangianNXproof} can be written as 
\begin{equation}
	{\cal L} = \frac{1}{f^{4 {\cal N}}} \int d^{4 \mathcal{N}} \theta  \, \Lambda^{2{\cal N}}\bar{\Lambda}^{2 {\cal N}}   
	\left( \mathbb{F} \, \bar{ \mathbb{F}} + f \, \mathbb{F}  + f \, \bar{\mathbb{F}}   \right)  
	\det \mathbb{A}_m^a  .
\end{equation}
Integrating now over superspace one finds
\begin{equation}
	\label{LGXF}
	{\cal L} = \left( \mathbb{F} \, \bar{ \mathbb{F}} +  f \, \mathbb{F}  +  f \, \bar{\mathbb{F}}   \right)  
	\det \mathbb{A}_m^a  |_{\theta^I=0} \, , 
\end{equation} 
giving the following equations of motion for the complex scalar $\mathbb{F}|_{\theta^I=0}$ 
\begin{equation}
	\mathbb{F}|_{\theta^I=0} = - f .
\end{equation} 
Substuting back this expression in the Lagrangian, \eqref{LGXF} reduces to \eqref{LNgeom}.

We conclude by observing that the Lagrangian \eqref{LGXF} contains the goldstini $\lambda_{I\alpha}$ inside $\det \mathbb{A}_m^a |_{\theta^I=0}$, but also a non-dynamical complex scalar field $ \mathbb{F}|_{\theta^I=0}$, transforming as a standard realization of the non-linear supersymmetry.
The presence of such non-dynamical complex scalar degree of freedom in the theory \eqref{lagrangianNX} is expected, since we know that in the linear realizations, beside the goldstini component fields $g_{I\alpha}$, there is an auxiliary component field $F$, which eventually is integrated out.

\end{document}